\documentclass{IEEEoj}
\usepackage{cite}
\usepackage{amsmath,amssymb,amsfonts}
\usepackage{graphicx,color}
\usepackage{textcomp}
\def\BibTeX{{\rm B\kern-.05em{\sc i\kern-.025em b}\kern-.08em
    T\kern-.1667em\lower.7ex\hbox{E}\kern-.125emX}}
\AtBeginDocument{\definecolor{ojcolor}{cmyk}{0.93,0.59,0.15,0.02}}
\def\OJlogo{\vspace{-4pt}\hskip-4pt\includegraphics[height=18pt]{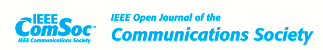}}
\IEEEoverridecommandlockouts
\usepackage{xspace,amsmath,amssymb,amsfonts,epsfig,subfigure,syntonly}
\usepackage{float}
\usepackage{ulem}
\usepackage{cancel}
\usepackage{soul}
\usepackage{amssymb}
\usepackage{lipsum}
\usepackage{cite,bm,color,url,textcomp,empheq,boxedminipage}
\usepackage{algorithmicx}
\usepackage{indentfirst}
\usepackage{epstopdf,makecell}
\usepackage{empheq}
\usepackage{pifont}
\usepackage{stfloats}
\usepackage[noend]{algpseudocode}
\usepackage{tcolorbox}
\usepackage[linesnumbered,ruled]{algorithm2e}
\usepackage{graphicx,graphics}  % Written by David Carlisle and Sebastian
\usepackage{multirow,multicol}
\usepackage{psfrag}    % Written by Craig Barratt, Michael C. Grant,
\usepackage{stfloats}
\usepackage{url}
\usepackage{lipsum}
\usepackage{lettrine}

\makeatletter

\newcommand{\Rmnum}[1]{\expandafter\@slowromancap\romannumeral #1@}
\makeatother

\long\def\symbolfootnote[#1]#2{\begingroup
\def\thefootnote{\fnsymbol{footnote}}
\footnote[#1]{#2}\endgroup}
\psfull

\makeatletter
\def\@invitedpaper{%
  \vspace{-0.8em}%
  \begin{center}%
    \textit{(Invited Paper)}%
  \end{center}%
  \vspace{0.5em}%
}
\makeatother

\begin{document}
\receiveddate{XX Month, XXXX}
\reviseddate{XX Month, XXXX}
\accepteddate{XX Month, XXXX}
\publisheddate{XX Month, XXXX}
\currentdate{2 June, 2025}
\doiinfo{OJCOMS.2024.011100}

\title{Joint User Association and Beamforming Design for ISAC Networks with Large Language Models}
\author{ Haoyun Li\IEEEauthorrefmark{1} \IEEEmembership{(Student Member, IEEE)}, Ming Xiao\IEEEauthorrefmark{1} \IEEEmembership{(Senior Member, IEEE)}, Kezhi Wang\IEEEauthorrefmark{2}  \IEEEmembership{(Senior Member, IEEE)}, Robert Schober\IEEEauthorrefmark{3} \IEEEmembership{(Fellow, IEEE)}, Dong In Kim\IEEEauthorrefmark{4}  \IEEEmembership{(Life Fellow, IEEE)},}
\author{Yong Liang Guan\IEEEauthorrefmark{5} \IEEEmembership{(Senior Member, IEEE)} \\ \textit{(Invited Paper)}}
\affil{Division of Information Science and Engineering, KTH Royal Institute of Technology, Stockholm 10044, Sweden.}

\affil{Department of Computer Science, Brunel University of
London, Uxbridge, Middlesex, UB8 3PH, UK.}

\affil{Institute for Digital Communications, Friedrich-Alexander-University Erlangen-Nurnberg (FAU), 91054 Erlangen, Germany.}

\affil{Department of Electrical and Computer Engineering, Sungkyunkwan University, Suwon 16419, South Korea.}

\affil{Continental-NTU Corporate Lab,
Nanyang Technological University, Singapore 639798.}

\authornote{CORRESPONDING AUTHOR: M. XIAO (e-mail: mingx@kth.se)}

\markboth{Joint User Association and Beamforming Design for ISAC Networks with Large Language Models}{H. Li \textit{et al.}}
\begin{abstract}
    Integrated sensing and communication (ISAC) has been envisioned to play a more important role in future wireless networks. However, the design of ISAC networks is challenging, especially when there are multiple communication and sensing (C\&S) nodes and multiple sensing targets. We investigate a multi-base station (BS) ISAC network in which multiple BSs equipped with multiple antennas simultaneously provide C\&S services for multiple ground communication users (CUs) and targets. To enhance the overall performance of C\&S, we formulate a joint user association (UA) and multi-BS transmit beamforming optimization problem with the objective of maximizing the total sum rate of all CUs while ensuring both the minimum target detection and parameter estimation requirements in terms of the radar signal-to-noise ratio (SNR) and the Cramér-Rao bound (CRB), respectively. To efficiently solve the highly non-convex mixed integer nonlinear programming (MINLP) optimization problem, we propose an alternating optimization (AO)-based algorithm that decomposes the problem into two sub-problems, i.e., UA optimization and multi-BS transmit beamforming optimization. Inspired by the huge potential of large language models (LLMs) for prediction and inference, we propose a unified framework integrating LLMs with convex-based optimization methods to benefit from the theoretical rigorousness and convergence guarantees of convex-based methods, and the adaptability and flexibility of LLMs. First, we propose a comprehensive design of prompt engineering based on in-context, few-shot, chain of thought, and self-reflection techniques to guide LLMs in solving the binary integer programming UA optimization problem. Second, we utilize convex-based optimization methods to handle the non-convex beamforming optimization problem based on fractional programming (FP), majorization minimization (MM), and the alternating direction method of multipliers (ADMM) with an optimized UA from LLMs. Numerical results demonstrate that our proposed LLM-enabled AO-based algorithm achieves fast convergence and near upper-bound performance with the GPT-o1 model, outperforming various benchmark schemes, which shows the advantages of integrating LLMs into convex-based optimization for wireless networks.
\end{abstract}
\maketitle

\section{Introduction}
The forthcoming sixth-generation (6G) wireless networks are expected to not only provide ubiquitous communication services with ultra-high data rates but also perform high-precision sensing \cite{isac_survey0, survey0}. The resulting communication and sensing (C\&S) system requires more resources in terms of frequency bands and hardware complexity to meet stringent multi-faceted performance requirements. Towards this end, integrated sensing and communication (ISAC) has been identified as a promising technology combining sensing and communication functionalities into a single system, in which C\&S share the same frequency bands and hardware platforms to substantially improve spectral and energy efficiency with limited hardware complexity \cite{isac_survey1, isac0, isac01}. Meanwhile, multi-input and multi-output (MIMO) techniques, first introduced in \cite{MIMO0}, have become a critical enabler for ISAC systems thanks to the large numbers of degrees of freedom (DoFs) offered by the multiple antennas deployed at both transmitters and receivers \cite{MIMO1, MIMO00}. With MIMO, ISAC systems can flexibly design transceiver beamforming and signal waveforms to generate directional beams towards the locations of CUs and sensing targets, thus enhancing communication rates and sensing accuracy \cite{Wang}.

Motivated by MIMO-enabled ISAC, there are a number of research works focusing on transmit beamforming design, including \cite{MIMO2, MIMO2_1, MIMO3,MIMO4,MIMO5,rangliu,MIMO6,MIMO7,MIMO8}. Specifically, the authors in \cite{MIMO2} considered an unmanned aerial vehicle (UAV)-enabled ISAC system, in which the UAV is equipped with a vertical uniform linear array (ULA) and acts as an aerial dual-functional access point (AP). The maneuvers and transmit beamforming of the UAV are jointly optimized to maximize the weighted sum rate of the CUs while ensuring the sensing requirements. As a further advance, the authors in \cite{MIMO2_1} aimed to jointly design the transmit beamforming and trajectories of UAVs equipped with ULAs in an ISAC network to maximize the achievable cover rate for legitimate ground users. To efficiently explore the spatial DoFs of the MIMO ISAC system, the authors in \cite{MIMO3} considered two different approaches to MIMO antenna usage: separate antenna deployment by partitioning MIMO antennas into radar and communication antennas, and sharing all the antennas for both radar sensing and downlink communications. The numerical results showed that the shared deployment achieved better performance in terms of the trade-off between C\&S. Different from \cite{MIMO3}, the authors in \cite{MIMO4} designed the transmit beamforming to fully extend the DoFs of the MIMO antennas through jointly utilizing the precoded individual communication and radar waveforms, which mitigates the interference between the communication and sensing signals. To improve the C\&S trade-off performance and its impact on cell-free (CF) MIMO systems, the authors in \cite{MIMO5} optimized transmit beamforming designs to suppress the interference and provide high-quality communication and sensing services, which were shown to outperform the zero-forcing (ZF) and the maximum ratio transmission (MRT) schemes. In \cite{rangliu}, the authors investigated a reconfigurable intelligent surface (RIS)-aided multi-user-MIMO (MU-MIMO) ISAC system, where the RIS reflection coefficients and transmit beamforming were jointly optimized to both enhance the target detection and parameter estimation performance with respect to the radar output signal-to-noise ratio (SNR) and the estimation Cramér-Rao bound (CRB), respectively.  Furthermore, the authors in \cite{MIMO6} considered an active RIS-empowered ISAC system, in which the transmit beamforming and active RIS beamforming were jointly optimized to minimize the CRB for target direct-of-arrival (DoA) estimation. To explore ISAC in vehicle-to-infrastructure (V2I) networks, the authors in \cite{MIMO7,MIMO8} proposed a deep learning (DL)-based approach to provide predictive beamforming for tracking on-road vehicles while reducing the signaling overhead and enhancing the ISAC performance.

Despite significant progress in MIMO ISAC system design, the existing results in \cite{MIMO2, MIMO2_1, MIMO3,MIMO4,MIMO5,rangliu,MIMO6,MIMO7,MIMO8} still face formidable challenges. The simultaneous execution of C\&S tasks in a MIMO ISAC system necessitates the joint design of transmit beamforming, transmit power allocation, and interference management strategies to achieve the expected trade-off between the C\&S performance metrics. The balance required among multiple objectives, such as sum rates,  target detection, and parameter estimation, complicates the system design and makes the underlying optimization problem non-trivial and computationally intensive.  Therefore, optimization tools, including model-based convex optimization \cite{MIMO2, MIMO2_1, MIMO3,MIMO4,MIMO5,rangliu,MIMO6}, data-driven DL \cite{MIMO7,MIMO8, Jiang}, and reinforcement learning (RL) \cite{RL1, RL2, RL3} are exploited to tackle the resulting highly complex optimization problems. As for convex-based optimization tools, such as the Lagrangian algorithm, successive convex approximation (SCA), and semidefinite relaxation (SDR) \cite{SDR}, the resulting optimization algorithms can generate a sub-optimal or even near-optimal solution with relatively low computational complexity, even for highly non-convex problems. The powerful data-driven capabilities of DL have advantages for model-free and non-linear mapping, as they are able to handle very high-dimensional input space while potentially learning to capture the characteristics of complex optimization problems \cite{DL}. RL-based optimization algorithms can learn the optimal policy by interacting with the system in real time, which is suitable for use cases with unknown system dynamics and complex reward functions \cite{RL1}. However, the above-mentioned tools have their limitations. To be more specific, as the network system complexity increases, convex-based optimization methods may suffer from the growing dimensionality and may get trapped in local convergence, while DL methods may be restricted by the growing complexity and volume of training data, computational demands and over-fitting risks, and for RL methods may have the issue of convergence and stability decrease \cite{MOPsurvey}. Another limitation of convex-based optimization, DL, and RL methods is their inherent reliance on expert domain knowledge. For instance, model-based convex optimization tools require expert knowledge, including communication theory, signal processing, and optimization theory. Model-free data-driven DL and RL methods demand network expert knowledge to formulate network optimization problems as regression tasks by correctly capturing network characteristics and mapping network status (inputs) to the policies (outputs) based on specifically designed neural network structures and the parameters learned from the labeled dataset \cite{Sun}. Both model-based and model-free methods require substantial expertise for each new network scenario, which can be time- and resource-consuming and substantially restrict scalability and flexibility in the dynamically changing environments of wireless networks. With superior generalization capabilities and knowledge integration, large language models (LLMs) can address this limitation by utilizing substantial inherent expert knowledge, thereby significantly reducing the expertise dependence and enhancing model adaptability and flexibility.

Recently, LLMs have emerged as powerful tools with demonstrated capabilities in natural language processing (NLP) applications, particularly for different aspects of problem optimization \cite{LLM01,LLM02,LLM03,LLM05,LLM06,LLM07}. For instance, the authors in \cite{LLM01, LLM02} leveraged LLMs as optimizers to solve general math problems via optimization task-specific prompts. The authors in \cite{LLM03} investigated a novel approach that integrated LLMs as black-box search operators into multi-objective evolutionary algorithms (MOEAs) in a zero-shot manner. The authors in \cite{LLM05} introduced a novel framework named LLAMBO that exploited the reasoning capabilities of LLM to enhance model-based Bayesian optimization via prompt engineering without fine-tuning. Moreover, the authors in \cite{LLM06, LLM07} explored the capabilities of LLMs in combinatorial optimization by integrating LLMs into the evolution of a  heuristics framework through problem-specific descriptions in prompts. LLMs offer several key advantages over traditional optimization approaches. First, LLMs exhibit strong adaptability across diverse tasks due to their pre-training on extensive datasets, enabling them to develop comprehensive domain knowledge \cite{LLMsurvey}. Unlike convex-based optimization methods, which require strict mathematical formulations, LLMs can handle general optimization problems using more natural problem representations. Additionally, LLMs can effectively leverage transfer learning for rapid domain adaptation with minimal fine-tuning requirements. Their ability to process natural language descriptions of optimization problems provides interpretability and flexibility advantages over conventional DL and RL approaches \cite{LLM22}. Consequently, recent studies have explored the potential of LLMs for wireless networks \cite{WirelessLLM,LLM4,LLM5,LLM6,LLM7,LLM8,LLM9}. For instance, the authors in \cite{WirelessLLM} proposed a novel framework named WirelessLLM empowered with knowledge and expertise for adapting and enhancing LLMs to solve general optimization problems in practical wireless scenarios like power allocation and spectrum sensing. The authors in \cite{LLM5, LLM6} leveraged LLMs for simple radio map generation and optimization for network planning, which improved the efficiency of AP deployment and management. By integrating LLMs as black-box search operators in evolutionary algorithms, the authors in \cite{LLM9} proposed a novel LLM-enabled multi-object evolutionary algorithm for finding the C\&S trade-off in a multi-UAV ISAC system, which outperformed traditional evolutionary algorithms in terms of the Pareto fronts and convergence obtained. All of the above works have demonstrated the capabilities of LLMs in solving optimization problems in wireless networks. Despite the promising potential of LLMs demonstrated in prior works \cite{WirelessLLM,LLM4,LLM5,LLM6,LLM7,LLM8,LLM9}, current LLMs still cannot efficiently solve highly complex non-convex optimization problems based only on prompts. More specifically, model-based and model-free optimization algorithms exploit gradient information of the objective function to precisely regulate the search direction and the step size at each iteration, which results in algorithm convergence. However, LLMs depend on discrete tokens from the prompt without information of numerical gradients. They lack the precision and stability for high-dimensional complex-valued computations and are unable to guarantee algorithm convergence due to the stochastic sampling of the iteratively generated tokens. Furthermore, LLMs are limited by the size of their context windows, which restricts them to articulating algorithmic concepts or pseudocodes rather than executing high-dimensional numerical optimization. Consequently, LLMs are more suitable as black-box heuristic solvers for optimization problems for which gradient information is not available. These observations motivate us to integrate LLMs with the existing optimization algorithm framework.

\begin{figure*}[!t]
    \centering
    \includegraphics[width=1\linewidth]{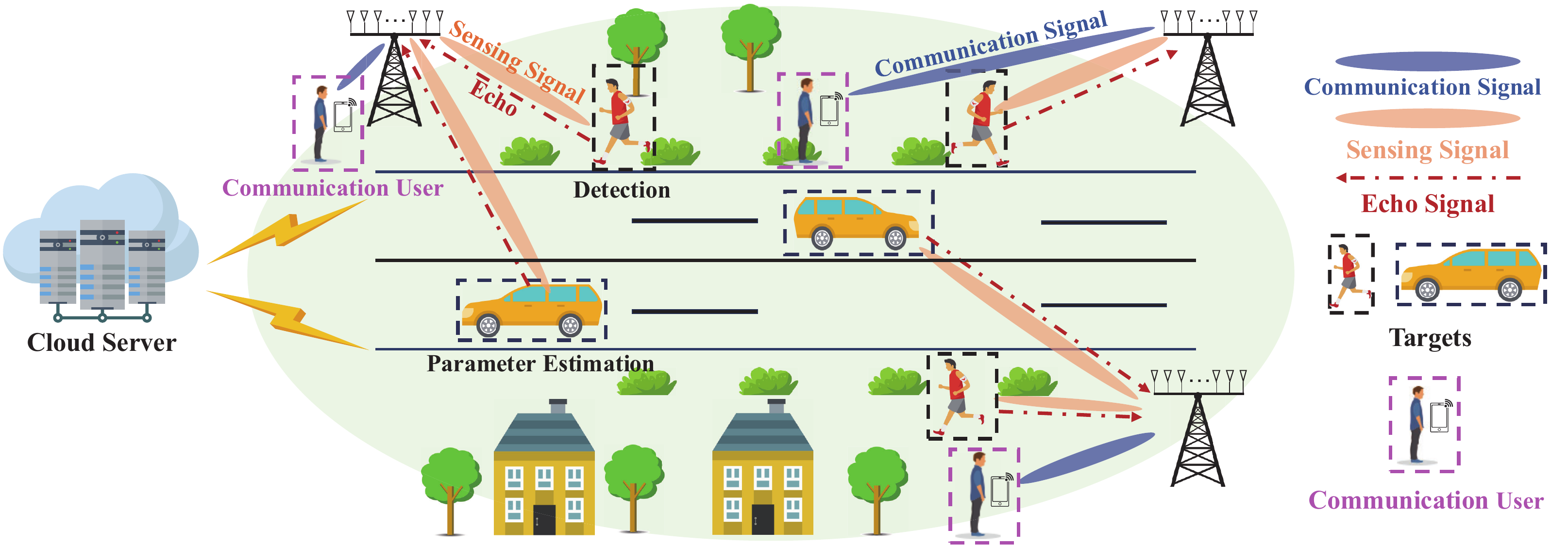}
    \caption{Illustration of the proposed multi-BS ISAC system.}
    \label{fig:system}
\end{figure*}

To the best of our knowledge, the integration of LLMs and convex-based optimization techniques for wireless communication network design remains unexplored. A unified framework that seamlessly integrates LLM-based techniques with traditional convex-based optimization could leverage the best of both worlds: the theoretical rigorousness and global convergence guarantees of convex-based optimization combined with the adaptability and flexibility of LLMs. Such a framework can provide the benefits of reducing extensive model training to adapt to wireless environments. Incorporating LLMs into the convex-based optimization framework can significantly alleviate the extensive reliance on expert-dependent training processes associated with traditional data-driven methods, which often require large-scale labeled datasets and domain-specific feature engineering, requiring considerable manual effort and expertise. In contrast, LLMs inherently capture generalized knowledge, which enables rapid adaptation to new optimization scenarios with minimal fine-tuning, reducing the need for explicit domain expertise and leading to shorter development cycles and lower training overhead.  Motivated by the above discussion, we propose a framework that integrates LLMs and convex-based optimization for a general MU-MIMO ISAC system, as shown in Fig. \ref{fig:system}, where multiple ISAC BSs serve multiple ground CUs and perform radar sensing for multiple targets. In particular, each BS is equipped with multiple antennas and simultaneously detects the presence of targets and estimates the DoAs, while all BSs collaboratively serve all the ground CUs. Under this setup, we aim to design the user association (UA) strategy jointly with the multi-BS transmit beamforming for optimization of the total communication performance of all CUs while ensuring the detection and DoA estimation requirements of multiple targets. The main contributions of this paper are summarized as follows:
\begin{itemize}
    \item Firstly, we investigate a multi-BS ISAC network, where multiple BSs equipped with multiple antennas simultaneously provide communication services for multiple ground CUs while performing sensing for multiple targets. To evaluate the C\&S performance, for radar sensing, we adopt the radar SNR as the target detection metric and the CRB as the target DoA estimation metric. For communications, we use the sum rate of all ground CUs as our communication metric. Based on the proposed metrics, we jointly take the UA strategy and the multi-BS transmit beamforming design into consideration to formulate the sum rate maximization problem while ensuring the detection and DoA estimation requirements of multiple targets.
    
    \item Secondly, to efficiently solve the highly non-convex problem, we first decompose the original problem into a UA sub-problem and a transmit beamforming sub-problem and solve them in an alternating optimization (AO) manner iteratively until convergence. We propose integrating LLMs and convex-based optimization into a single framework to exploit the theoretical rigorousness and convergence guarantees of convex-based optimization with the adaptability and flexibility of LLMs. Specifically, for the integer programming UA optimization problem, we propose a novel LLM-based algorithm based on prompt engineering to generate a high-quality UA strategy. To effectively adapt LLMs for optimization, we specifically design the prompt engineering based on the in-context, few-shot, chain of thought, and self-reflection prompting design techniques. For the transmit beamforming optimization problem, we decompose it into a series of sub-problems with respect to the transmit beamforming at each BS and solve them simultaneously for a given UA. However, the resulting sub-problem is still highly non-convex due to the objective function and the sensing requirement constraints. To tackle this issue, we propose an alternating direction method of multipliers (ADMM)-based algorithm using the techniques of fractional programming (FP), augmented Lagrangian algorithm, and majorization-minimization (MM) to iteratively optimize the transmit beamforming for each BS.
    
    \item Thirdly, we conduct extensive simulations to show the effectiveness of the proposed LLM-enabled AO-based algorithm design compared with several benchmark methods. We also compare the performance of multiple LLMs, including GPT-o1, GPT-4-Turbo, Claude 3.5, and Gemini 2.0. It is shown that the proposed algorithm using the GPT-o1 model can achieve a similar, and sometimes the same performance as an upper bound and outperforms a benchmark method using only convex-based optimization techniques or other LLM-based schemes in terms of the total sum rate and convergence speed. This demonstrates the effectiveness of our designed prompt engineering for adapting LLMs to the optimization problem. In addition, with increasing system complexity, the proposed algorithm can still approach the upper bound and outperform benchmark schemes, which reveals the robustness and efficiency of the proposed algorithm.
\end{itemize}

The remainder of this paper is organized as follows. Section II presents the system model of the considered multi-CU multi-BS ISAC system. Section III formulates the optimization problem for total sum rate maximization while ensuring the detection and DoA estimation requirements of multiple targets. Section IV introduces the LLM-enabled AO-based algorithm design for solving the formulated problem. Section V provides numerical results to study the effectiveness of the proposed algorithm. Finally, Section VI concludes the paper.

\textit{Notations}: Unless otherwise specified, we use boldface lowercase letters to denote vectors and boldface uppercase letters to denote matrices. $(\cdot)^H$, $(\cdot)^T$, and $\bar{(\cdot)}$ denote the conjugate transpose, transpose, and conjugate, respectively. For a square matrix $\mathbf{A}$, $\text{Tr}\{\mathbf{A}\}$ and $\mathbf{A}^{-1}$ denote its trace and inverse, respectively, and $\mathbf{A} \succeq \mathbf{0}$ indicates that $\mathbf{A}$ is positive semidefinite. $[\mathbf{A}]_{n,p}$ denotes the $(n,p)$-th entry of $\mathbf{A}$. $\mathbb{C}^{M \times N}$ denotes the set of $M \times N$ complex matrices. $\mathbf{I}_N$ denotes the $N \times N$ identity matrix. $\mathcal{CN}(\boldsymbol{\mu}, \boldsymbol{\Sigma})$ denotes the circularly symmetric complex Gaussian (CSCG) distribution, where $\boldsymbol{\mu}$ and $\boldsymbol{\Sigma}$ are the mean vector and the covariance matrix, respectively. $\mathbb{E}\{\cdot\}$ denotes the expectation operator. $||\cdot||_2$ denotes the Euclidean norm of a complex vector. $||\cdot||_F$ denotes the Frobenius norm of a complex matrix. $||\cdot||_0$ denotes the L0 norm of a vector. $|\cdot|$ denotes the magnitude of a complex number. $\frac{\partial f(x,y,\cdots)}{\partial x}$ denotes the  partial derivative of a function $f(x,y,\cdots)$ with respect to variable $x$. $\otimes$ denotes the Kronecker product. $\mathfrak{j}$ denotes the imaginary unit. In addition, $\text{Re}\{\cdot\}$ and $\text{Im}\{\cdot\}$ denote the real and imaginary parts of a complex-valued matrix, respectively.

\section{System Model}
We consider a multi-BS ISAC system, as shown in Fig. \ref{fig:system}, where $K$ ISAC BSs serve a total of $N$ single-antenna ground CUs and perform radar sensing for a total of $Q$ targets. We denote the set of ISAC BSs as $\mathcal{K}=\{1,2,\dots,K\}$, the set of ground CUs as $\mathcal{N}=\{1,2, \dots, N\}$, and the set of targets as $\mathcal{Q}=\{\mathcal{Q}_1,\mathcal{Q}_2,\dots,\mathcal{Q}_K\}$, respectively. In this system, each ISAC BS is assumed to be a dual-functional radar-communication (DFRC) BS equipped with $M$ antennas. We assume that all ISAC BSs can receive sensing echo signals and maintain downlink communications concurrently without interference by exploiting full-duplex radio technologies \cite{isac1,isac2}. All ISAC BSs are supposed to serve all ground CUs collaboratively. We assume that ISAC BS $k$ senses the set of targets in the area of interest, denoted as $\mathcal{Q}_k$, and $\mathcal{Q}_1 \cup \mathcal{Q}_2 \cup \cdots \cup \mathcal{Q}_K=\mathcal{Q}$. To simplify our system, we model all targets as point targets. In this system, we assume that each ISAC BS employs a different orthogonal frequency band to eliminate inter-BS interference \cite{Zhao}. 
\subsection{Signal Model}
Let $s^c_{k,i}[l] \in \mathbb{C}$ denote the complex communication symbol transmitted from BS $k$ to CU $i$ in time slot $l$, and $\mathbf{W}^c_k = \left[\mathbf{w}^c_{k,1}, \mathbf{w}^c_{k,2}, \cdots, \mathbf{w}^c_{k,N} \right] \in \mathbb{C}^{M \times N}$ denote the transmit communication beamforming matrix from BS $k$ to all CUs, where $\mathbf{w}^c_{k,i} \in \mathbb{C}^{M \times 1}$ denotes the dedicated communication beamforming vector for the $i$-th CU. Hence, the transmitted communication signal is given by
\begin{equation}
    \mathbf{x}^c_k[l] = \mathbf{W}^c_k \mathbf{s}^c_k[l] \in \mathbb{C}^{M \times 1},
\end{equation}
where $\mathbf{s}^c_k[l]=[s^c_{k,1}[l], s^c_{k,2}[l], \cdots, s^c_{k,N}[l]]^T \in \mathbb{C}^{N \times 1}$ denotes the transmitted communication symbols from BS $k$ to all CUs in time slot $l$. It is assumed that $s^c_{k,i}[l], \forall k, \forall i$ are mutually independent with zero mean and unit energy, where $\mathbb{E}\{|s^c_{k, i}[l]|^2\}=1, \mathbb{E}\left\{\mathbf{s}^c_k[l]\mathbf{s}^{c}_{k}[l]^H\!\right\} = \mathbf{I}_N$. Similarly, the complex radar signal transmitted by BS $k$ in time slot $l$ can be expressed as
\begin{equation}
    \mathbf{x}^r_k[l] = \mathbf{W}^r_k \mathbf{s}^r_k[l] \in \mathbb{C}^{M \times 1},
\end{equation}
where $\mathbf{W}^r_k \in \mathbb{C}^{M \times M}$ denotes the transmit radar beamforming matrix of BS $k$, and $\mathbf{s}^r_k[l] \in \mathbb{C}^{M \times 1}$ denotes the transmitted radar signal of BS $k$ for the targets in $\mathcal{Q}_k$, which includes $M$ individual independent waveforms with $\mathbb{E}\left\{\mathbf{s}^r_k[l]\mathbf{s}^{r}_{k}[l]^H\!\right\} = \mathbf{I}_M$. We assume the communication symbols are independent from the radar waveforms with $\mathbb{E}\left\{\mathbf{s}^c_k[l]\mathbf{s}^{r}_{k'}[l]^H\!\right\} = \mathbf{0}, ~\forall k, k' \in \mathcal{K}, ~k \neq k'$.

The ISAC signal transmitted by BS $k$ in time slot $l$ is given by
\begin{align}
        \mathbf{x}_k[l]&=\mathbf{x}^c_{k}[l]+\mathbf{x}^r_{k}[l] \\ \notag
        &=\mathbf{W}^c_k\mathbf{s}^c_k[l]+\mathbf{W}^r_k\mathbf{s}^r_k[l]\\ \notag
        &=\mathbf{W}_k\mathbf{s}_k[l] \in \mathbb{C}^{M \times 1}, 
\end{align}
where $\mathbf{W}_k=[\mathbf{W}^c_k,\mathbf{W}^r_k] \in \mathbb{C}^{M \times (N+M)}$ denotes the downlink transmit beamforming matrix of BS $k$, and $\mathbf{s}_k[l]=\left[\mathbf{s}^c_k[l]^T, \mathbf{s}^r_k[l]^T\right]^T \in \mathbb{C}^{(N+M)\times 1}$ denotes the transmitted symbol vector of BS $k$ in time slot $l$.
\subsection{Communication Model}
We denote $\mathbf{h}_{k,i} \in \mathbb{C}^{M \times 1}$ as the communication channel between BS $k$ and CU $i$, which is assumed to follow Rician fading and given by
\begin{equation}
    \mathbf{h}_{k,i}=\underbrace{\beta_{k,i} \sqrt{\frac{\kappa_{k,i}}{\kappa_{k,i}+1}} \overline{\mathbf{c}}_{k,i}}_{\triangleq \overline{\mathbf{h}}_{k,i}}+\underbrace{\beta_{k,i} \sqrt{\frac{1}{\kappa_{k,i}+1}} \tilde{\mathbf{c}}_{k, i}}_{\triangleq \tilde{\mathbf{h}}_{k, i}}.
\end{equation}
Here, $\beta_{k,i}=\beta_0(\frac{d_{k,i}}{d_0})^{-\varsigma_c}$ is the large-scale fading factor from BS $k$ to CU $i$, where $\beta_0$ is the path loss factor at a reference distance $d_0=1$ m, $d_{k,i}$ is the distance between BS $k$ and CU $i$, and $\varsigma_c$ is the path loss exponent.  $\kappa_{k,i}$ denotes the Rician factor of the channel from BS $k$ to CU $i$. $\tilde{\mathbf{c}}_{k, i}$ is the non-line-of-sight (NLoS) component, where elements are an independent complex CSCG random variables with zero mean and unit variance, that is $\tilde{\mathbf{c}}_{k, i} \sim {\mathcal{CN}}\left(0, \mathbf{I}_M\right)$, and $\overline{\mathbf{c}}_{k,i}$ is the line-of-sight (LoS) component given as $\overline{\mathbf{c}}_{k,i} =\frac{1}{\sqrt{M}}\boldsymbol{a}\left(\theta_{k,i}\right)$, where $\boldsymbol{a}\left(\theta_{k,i}\right)$ is the steering vector, and $\theta_{k,i}$ is the direction of departure (DoD) from BS $k$ to CU $i$. We assume a ULA at each BS, and $\boldsymbol{a}\left(\theta_{k,i}\right)$ can be expressed as
\begin{equation}
    \boldsymbol{a}\left(\theta_{k,i}\right) = \left[\begin{array}{lll}
1,e^{j \frac{2\pi d}{\lambda} \cos \left(\theta_{k,i}\right)},\cdots,e^{(M-1) j \frac{2\pi d}{\lambda} \cos \left(\theta_{k,i}\right)}
\end{array}\right]^T,
\end{equation}
where $d$ denotes the antenna spacing distance, and $\lambda$ is the wavelength.
Then, the signal received at CU $i$ from BS $k$ can be expressed as
\begin{align}
    y^c_{k,i}[l]&=\mathbf{h}^H_{k,i}\mathbf{x}_k[l]+z_i[l]\\\notag
    &=\mathbf{h}^H_{k,i}\mathbf{w}^c_{k,i}s^c_{k,i}[l]+\mathbf{h}^H_{k,i}\sum\limits_{n\in\mathcal{N},n\neq i}\mathbf{w}^c_{k,n}s^c_{k,n}[l]\notag\\
    &+ \mathbf{h}^H_{k,i}\mathbf{W}^r_k\mathbf{s}^r_k[l] + z_i[l],
\end{align}
where $z_i[l]$ denotes the complex circularly symmetric white Gaussian noise received at CU $i$ with a zero mean and variance $\sigma^2_i$. We assume that each ISAC BS employs a different orthogonal frequency band to eliminate inter-BS interference \cite{Zhao}. Therefore, the signal-to-interference-plus-noise ratio (SINR) from BS $k$ to CU $i$ is given by
\begin{equation}
    \gamma_{k,i}=\frac{|\mathbf{h}^H_{k,i}\mathbf{w}_{k,i}|^2}{\sum_{n=1, n \neq i}^{N+M}|\mathbf{h}^H_{k,i}\mathbf{w}_{k,n}|^2+\sigma^2_i},
\end{equation}
where $\mathbf{w}_{k,i}$ denotes the $i$-th column of $\mathbf{W}_k=[\mathbf{w}_{k,1}, \mathbf{w}_{k,2}, \cdots, \mathbf{w}_{k,N+M}]$.

\subsection{Sensing Model}
We assume that each BS $k$ is supposed to sense a specific subset of targets in the area of interest of BS $k$, i.e., $\mathcal{Q}_k$, and there is no overlap between any two target subsets, i.e., $\mathcal{Q}_k \cap \mathcal{Q}_j=\emptyset, ~\forall k,j \in \mathcal{K}, ~k \neq j$, and $\mathcal{Q}_1\cup\mathcal{Q}_2\cdots\cup\mathcal{Q}_K=\mathcal{Q}$. In each subset $\mathcal{Q}_k$, there is one target $\tilde{Q}_k$ for detection and one target $\hat{Q}_k$ for parameter estimation, respectively. To simplify illustration, we assume all BSs operate in the time division multiple access (TDMA) mode or apply enhanced
inter-cell interference coordination techniques such as those in
LTE Rel.10 to limit inter-cell interference \cite{Astely}. The signal echoes transmitted by BS $k$ reflected by targets and received at BS $k$ in time slot $l$ are given by
\begin{equation}
    \mathbf{y}^r_k[l]=\sum\limits_{i\in\mathcal{Q}} \alpha_{t,i} \mathbf{g}_{k,i}(\phi_{k,i})\mathbf{g}^H_{k,i}(\phi_{k,i})\mathbf{W}_k\mathbf{s}_k[l]+\mathbf{z}_{k}[l], \label{sense_signal}
\end{equation}
where $\alpha_{t,i}\sim \mathcal{CN}(0,\sigma^2_t)$ denotes the radar cross section (RCS) of the $i$-th target, $\mathbf{g}_{k,i}\in \mathbb{C}^{M \times 1}$ denotes the sensing channel between BS $k$ and target $i$, and $\mathbf{z}_{k}[l]\sim \mathcal{CN}(0, \sigma^2_r\mathbf{I}_M)$ is the CSCG noise. It is assumed that the channels between the BSs and the targets are LoS. To be more specific, $\mathbf{g}_{k,i}=\frac{1}{\sqrt{M}}\tilde{\beta}_{k,i}\boldsymbol{a}(\phi_{k,i})$, where $\tilde{\beta}_{k,i}=\beta_0(\frac{d_{k,i}}{d_0})^{-\varsigma_t}$ is the large-scale fading factor from BS $k$ to target $i$, $\varsigma_t$ is the path loss exponent, and $\phi_{k,i}$ is the DoA with respect to target $i$ and BS $k$. 
% As a result, the signal echoes transmitted from BS $k$ reflected by targets $i \in \mathcal{Q}_k$ received at BS $k$ in time slot $l$ is given by
% \begin{equation}
%     \mathbf{y}^r_{k,i}[l]=\alpha_{t,i} \mathbf{g}_{k,i}(\phi_{k,i})\mathbf{g}^H_{k,i}(\phi_{k,i})\mathbf{W}_k\mathbf{s}_k[l]+\mathbf{z}_{k}[l], ~\forall i \in \mathcal{Q}_k. \label{sense_signal2}
% \end{equation}
\section{Performance Metrics}
In this section, we first separately derive the sensing detection and parameter estimation performance metrics, and then formulate the sum rate of all ground CUs as our communication metric.

\subsection{Target Detection}
In this section, we will show that the probability of target detection is positively related to the radar SNR, which is used as our sensing performance metric for target detection. Assume that the system has prior knowledge of the DoA of each target via some localization algorithms or estimation methods based on previous observations \cite{Bekkerman}. Each BS $k, ~\forall k \in \mathcal{K}$, implements a receive beamforming vector $\frac{1}{\sqrt{M}}\boldsymbol{a}^H(\dot{\phi}_{k,i})$ to multiply with the echo signal reflected by target $i$. The sensing signal reflected by target $i$ received at BS $k$ can be approximately expressed as
\begin{align}
    \tilde{y}^r_{k,i}[l]
    &=\alpha_{t,i}\tilde{\beta}_{k,i}\mathbf{g}^H_{k,i}(\phi_{k,i})\mathbf{W}_k{\mathbf{s}}_k[l]+z_{k}[l]. \label{eq:signal}
\end{align}
We denote $\Dot{\phi}_{k,i}$ as the estimate of $\phi_{k,i}$ and assume $\Dot{\phi}_{k,i}\approx\phi_{k,i}$ for simplification.  We have $\forall i \neq i', |\boldsymbol{a}^H({\phi}_{k,i})\boldsymbol{a}(\phi_{k,i'})|\approx0,\frac{1}{M}\boldsymbol{a}^H(\Dot{\phi}_{k,i})\boldsymbol{a}(\phi_{k,i})\approx1$ if $M$ is sufficiently large \cite{book3}. In addition, $z_{k}[l]=\frac{1}{\sqrt{M}}\boldsymbol{a}^H(\Dot{\phi}_{k,i})\mathbf{z}_{k}[l] \in \mathbb{C}$, is a white CSCG noise, i.e., $z_{k}[l]\sim \mathcal{CN}(0,\sigma^2_r)$. To improve the sensing accuracy, we collect samples of a total of $L$ time slots. By averaging over $L$ time slots, we can express the final received sensing samples as
\begin{align}
    \tilde{y}^r_{k,i}&=\frac{1}{L}\sum\limits_{l=1}^L\tilde{y}^r_{k,i}[l],\\ \notag
    &=\underbrace{\frac{1}{L}\sum\limits_{l=1}^L\alpha_{t,i}\tilde{\beta}_{k,i}\mathbf{g}^H_{k,i}(\phi_{k,i})\mathbf{W}_k{\mathbf{s}}_k[l]}_{\triangleq \eta_k}+\underbrace{\frac{1}{L}\sum\limits_{l=1}^Lz_{k}[l]}_{\triangleq \tilde{z}_k}.
\end{align}

Thus, the target detection problem can be formulated as a composite binary hypothesis test as follows:
\begin{equation}
y=\left\{\begin{array}{l}
\mathcal{H}_0: z_k, \\
\mathcal{H}_1: \eta_k+\tilde{z}_k,
\end{array}\right.
\end{equation}
where $\eta_k \triangleq \frac{1}{L}\sum\limits_{l=1}^L\alpha_{t,i}\tilde{\beta}_{k,i}\mathbf{g}^H_{k,i}(\phi_{k,i})\mathbf{W}_k{\mathbf{s}}_k[l]$, $\alpha_{t,i} \sim \mathcal{C} \mathcal{N}\left(0, \sigma_{\mathrm{t}}^2\right)$, and $\tilde{z}_k \triangleq \frac{1}{L}\sum\limits_{l=1}^Lz_{k}[l]$. The Neyman-Pearson detector is given by \cite{book2}
\begin{equation}
    E=|y|^2 \underset{\mathcal{H}_1}{\stackrel{\mathcal{H}_0}{\lessgtr}} \delta,
\end{equation}
where $\delta$ is the decision threshold determined by the probability of a false alarm. Noting that $\alpha_{t,i}$ is independent of $\mathbf{{s}}_k[l], \forall l$, we have the conditional probability distributions of $y \mid \mathcal{H}_0 \sim \mathcal{CN}(0, \mu_0)$ with $\mu_0\triangleq\mathbb{E}\{|\tilde{z}_k|^2\}=\frac{\sigma^2_r}{L}$ and $y \mid \mathcal{H}_1 \sim \mathcal{CN}(0, \mu_1)$ with $\mu_1\triangleq\sigma^2_t|\tilde{\beta}_{k,i}|^2\frac{\mathbf{W}^H_k\mathbf{g}_{k,i}\mathbf{g}^H_{k,i}\mathbf{W}_k}{L}+\frac{\sigma^2_r}{L}$. Therefore, the test statistic $E$ is subject to the following distribution \cite{book2}
\begin{equation}
    E \sim \begin{cases}\frac{\sigma^2_r}{L} \chi_2^2, & \mathcal{H}_0 \\ \left(\sigma^2_t|\tilde{\beta}_{k,i}|^2\frac{\mathbf{W}^H_k\mathbf{g}_{k,i}\mathbf{g}^H_{k,i}\mathbf{W}_k}{L}+\frac{\sigma^2_r}{L}\right) \chi_2^2, & \mathcal{H}_1\end{cases},
\end{equation}
where $\chi_2^2$ denotes the central chi-squared distribution with two DoFs. As a result, the probability of a false alarm can be calculated as
\begin{equation}
    \text{P}_{\text{FA}} = \text{Pr}(E>\delta|\mathcal{H}_0)=1-\text{F}_{\chi^2_2}\left(\delta/\mu_0\right),
\end{equation}
where $\text{Pr}(\cdot)$ denotes the probability, and $\text{F}_{\chi^2_2}(\cdot)$ is the cumulative distribution function (CDF) of the chi-square random variable. Similarly, the probability of detection can be calculated as
\begin{equation}
    \text{P}_{\text{D}} = \text{Pr}(E>\delta|\mathcal{H}_1)=1-\text{F}_{\chi^2_2}\left(\delta/\mu_1\right).
\end{equation}
For a given decision threshold $\delta$, the detection probability $\text{P}_\text{D}$ can be rewritten as 
\begin{equation}
    \text{P}_\text{D}=1-\text{F}_{\chi^2_2}\left(\frac{\mu_0}{\mu_1}\text{F}^{-1}_{\chi^2_2}(1-\text{F}_\text{FA})\right),
\end{equation}
where $\text{F}^{-1}_{\chi^2_2}(\cdot)$ denotes the inverse function of $\text{F}_{\chi^2_2}$. It is clearly observed that $\text{P}_\text{D}$ is positively related to $\frac{\mu_1}{\mu_0}$, which is also positively related to radar SNR $\Gamma_{k,i}$ as
\begin{equation}
    \Gamma_{k,i} = \frac{\sigma^2_t|\tilde{\beta}_{k,i}|^2\mathbf{W}^H_k\mathbf{g}_{k,i}\mathbf{g}^H_{k,i}\mathbf{W}_k}{\sigma^2_r}.
\end{equation}
Therefore, we use the radar SNR $\Gamma_{k,i}$ as the sensing performance metric for target detection.
\subsection{Parameter Estimation}
In this section, we derive the CRB of the DoA of target $i$ as our sensing performance metric for parameter estimation. To start with, we collect the unknow parameters in vector $\boldsymbol{\xi}_{k,i} \triangleq [\phi_{k,i}, \boldsymbol{\alpha}_{i}^T]^T \in \mathbb{C}^{3\times1}$ with $\boldsymbol{\alpha}_i\triangleq[\text{Re}(\alpha_{t,i}), \text{Im}(\alpha_{t,i})]^{\it{T}} \in \mathbb{C}^{2\times1}$, and we focus on DoA estimation of $\phi_{k,i}$. To improve the estimation accuracy, we collect samples of all $L$ time slots with $\tilde{\mathbf{y}}^r_{k}=\left[(\mathbf{y}^r_{k}[1])^T, (\mathbf{y}^r_{k}[2])^T, \cdots, (\mathbf{y}^r_{k}[L])^T\right]^T \in \mathbb{C}^{ML\times1}$. For simplification, we define $\mathbf{G}(\phi_{k,i})\triangleq\mathbf{g}(\phi_{k,i})\mathbf{g}^H(\phi_{k,i}) \in \mathbb{C}^{M\times M}$, $\tilde{\mathbf{z}}_k\triangleq\left[\mathbf{z}^T_k[1], \mathbf{z}^T_k[2], \cdots, \mathbf{z}^T_k[L]\right]^T \in \mathbb{C}^{ML\times1}$, \\
and
\begin{align}
   \boldsymbol{\mu} (\boldsymbol{\xi}_{k,i})\triangleq&\bigg[(\alpha_{t,i} \mathbf{G}(\phi_{k,i})\mathbf{W}_k\mathbf{s}_k[1])^T, \notag \\
&\cdots, (\alpha_{t,i} \mathbf{G}(\phi_{k,i})\mathbf{W}_k\mathbf{s}_k[L])^T\bigg]^T \in \mathbb{C}^{ML\times1} 
\end{align}
 which is defined according to \eqref{sense_signal}, respectively. Thus, we have
\begin{equation}
\tilde{\mathbf{y}}^r_{k}=\sum\limits_{i\in\mathcal{Q}}\boldsymbol{\mu}(\boldsymbol{\xi}_{k,i})+\tilde{\mathbf{z}}_k \in \mathbb{C}^{ML\times1},
\end{equation}
where $\tilde{\mathbf{y}}^r_{k} \sim \mathcal{CN}(\sum\nolimits_{i\in\mathcal{Q}}\boldsymbol{\mu}(\boldsymbol{\xi}_{k,i}), \sigma^2_r\mathbf{I}_{{ML}})$, $\mathbb{E}\{\tilde{\mathbf{z}}_k\tilde{\mathbf{z}}^H_k\}=\sigma^2_r\mathbf{I}_{{ML}}$. The $(n,p)$-th element of the Fisher Information Matrix (FIM) for estimating parameters $\boldsymbol{\xi}_{k,i}$ based on the model in \eqref{sense_signal} is given by \cite{Bekkerman}
\begin{equation}
    [\mathbf{J}(\boldsymbol{\xi}_{k,i})]_{n, p}=\frac{2}{\sigma_r^2} \operatorname{Re}\left\{\frac{\partial \boldsymbol{\mu}(\boldsymbol{\xi}_{k,i})^H}{\partial \xi_n} \frac{\partial \boldsymbol{\mu}(\boldsymbol{\xi}_{k,i})}{\partial \xi_p}\right\}, \label{FIM1}
\end{equation}
where $\xi_n$ denotes the $n$-th element in $\boldsymbol{\xi}_{k,i}$, and $\mathbf{J}(\boldsymbol{\xi}_{k,i}) \in \mathbb{C}^{3\times3}$ is the FIM, which can be partitioned into a block matrix given as
\begin{equation}
    \mathbf{J}(\boldsymbol{\xi}_{k,i})=\left[\begin{array}{ll}
{J}(\boldsymbol{\xi}_{k,i})_{\phi_{k,i}\phi_{k,i}} & \mathbf{J}(\boldsymbol{\xi}_{k,i})_{\phi_{k,i}\boldsymbol{\alpha}^T_{i}} \\
\mathbf{J}(\boldsymbol{\xi}_{k,i})^T_{\phi_{k,i}\boldsymbol{\alpha}^T_{i}} & \mathbf{J}(\boldsymbol{\xi}_{k,i})_{\boldsymbol{\alpha}_{i}\boldsymbol{\alpha}^T_{i}}
\end{array}\right], \label{FIM2}
\end{equation}
where the real ${J}(\boldsymbol{\xi}_{k,i})_{\phi_{k,i}\phi_{k,i}}$ and sub-matrices in $\mathbf{J}(\boldsymbol{\xi}_{k,i})$ are derived in Appendix \ref{appenA}. The CRB for DoA $\phi_{k,i}$ estimation of the target can be expressed as
\begin{align}
\begin{split}
\mathrm{CRB}(\phi_{k,i})&=\, \Biggl[
{J}(\boldsymbol{\xi}_{k,i})_{\phi_{k,i}\phi_{k,i}}\notag \\
&-\, \mathbf{J}(\boldsymbol{\xi}_{k,i})_{\phi_{k,i}\boldsymbol{\alpha}^T_{i}}
\mathbf{J}(\boldsymbol{\xi}_{k,i})^{-1}_{\boldsymbol{\alpha}_{i}\boldsymbol{\alpha}^T_{i}} \mathbf{J}(\boldsymbol{\xi}_{k,i})^{T}_{\phi_{k,i}\boldsymbol{\alpha}^T_{i}}\Biggr]^{-1}.
\end{split}
\end{align}

\subsection{Communication Metrics}
In this section, we derive the sum rate of all ground CUs as our communication performance metric. To start with, we define the UA matrix $\mathbf{U}\in\mathbb{C}^{K\times N}$ as
\begin{equation}
\mathbf{U}=\left[\mathbf{u}_1, \mathbf{u}_2, \cdots, \mathbf{u}_N\right]=\left[\begin{array}{c}
\underbrace{u_{1,1}, u_{1,2}, \cdots, u_{1, N}}_{\widetilde{\mathbf{u}}_1} \\
\vdots \\
\underbrace{u_{K, 1}, u_{K, 2}, \cdots, u_{K, N}}_{\widetilde{\mathbf{u}}_K}
\end{array}\right].
\end{equation}
We denote $\mathbf{u}_i=[u_{1,i}, u_{2,i}, \cdots, u_{K,i}]^T\in \mathbb{C}^{K\times 1}$ as the UA vector of CU $i$, where $u_{k,i}$ is a binary variable. Specifically, $u_{k,i}\in \{0,1\}, ~\forall k, ~\forall i$, and $u_{k,i}=1$ if CU $i$ is associated with BS $k$. Otherwise, $u_{k,i}=0$. $||\widetilde{\mathbf{u}}_k||_0$ is the number of CUs served by BS $k$. It is assumed that each CU can only be served by one BS, which means that $||\mathbf{u}_i||_0=1$. Therefore, the achievable data rate from BS $k$ to CU $i$ is defined as $R_{k,i}$, which is given by
\begin{equation}
    R_{k,i}=\frac{B}{||\widetilde{\mathbf{u}}_k||_0}\mathrm{log}_2(1+\gamma_{k,i}), \label{R_KI}
\end{equation}
where $B$ is the total bandwidth allocated to BS $k$.
Based on \eqref{R_KI}, the achievable data rate of CU $i$ in bps/Hz can be rewritten as
\begin{equation}
    R_i=\sum\limits_{k\in \mathcal{K}}u_{k,i}\frac{B}{||\widetilde{\mathbf{u}}_k||_0}\mathrm{log}_2(1+\gamma_{k,i}).
\end{equation}

As a result, the sum rate of all CUs can be expressed as
\begin{align}
    R&=\sum\limits_{i\in \mathcal{N}}R_i=\sum\limits_{i\in \mathcal{N}}\sum\limits_{k\in \mathcal{K}}u_{k,i}\frac{B}{||\widetilde{\mathbf{u}}_k||_0}\times \notag \\ &\mathrm{log}_2\left(1+\frac{|\mathbf{h}^H_{k,i}\mathbf{w}_{k,i}|^2}{\sum_{n=1,n \neq i}^{N+M}|\mathbf{h}^H_{k,i}\mathbf{w}_{k,n}|^2+\sigma^2_i}\right).
\end{align}

\section{Problem Formulation}
In this paper, we aim to maximize the sum rate of all ground CUs while guaranteeing the sensing performance of all targets subject to the maximum transmit power constraints and UA requirements. In this section, we first formulate the joint BS beamforming matrices and UA strategy optimization problem. Then, we decompose the original problem into UA and BS transmit beamforming matrix optimization sub-problems.

\subsection{ Formulation of Problem}
Based on the performance metrics derived in Section III, our goal is to maximize the total sum rate of all ground CUs subject to the radar SNR and DoA estimation CRB constraints of all targets and the maximum BS transmit power budget constraints by jointly optimizing with BS transmit beamforming matrices $\{\mathbf{W}_k\}^K_{k=1}$ and the UA strategy $\mathbf{U}$. Therefore, the optimization problem can be formulated as follows
\begin{subequations}\label{eq:main}
% 主公式部分
\begin{align}
&\text{(P1):}\quad \max_{\{\mathbf{W}_k\}_{k=1}^K,\, \mathbf{U}} 
\; \sum_{i\in \mathcal{N}} \sum_{k\in \mathcal{K}} u_{k,i}\,\frac{B}{\|\widetilde{\mathbf{u}}_k\|_0} \times\notag \\ & \, \qquad \log_2\Biggl(1+\frac{|\mathbf{h}^H_{k,i}\mathbf{w}_{k,i}|^2}
{\sum_{n=1,n\neq i}^{N+M} |\mathbf{h}^H_{k,i}\mathbf{w}_{k,n}|^2+\sigma^2_i}\Biggr)
\tag{\ref{eq:main}}\label{eq:main_obj}
\end{align}

% 约束部分
\begin{align}
~\text{s.t.}\quad
& \frac{\sigma^2_t|\tilde{\beta}_{k,i}|^2\,\mathbf{W}^H_k\,\mathbf{g}_{k,i}\mathbf{g}^H_{k,i}\,\mathbf{W}_k}{\sigma^2_r}
\geq \Gamma_\text{t}, \nonumber~\forall\, k\in \mathcal{K},\; i\in \tilde{\mathcal{Q}}_k,
\tag{\ref{eq:main}a}\label{eq:main_a}\\[1ex]
& \operatorname{CRB}(\phi_{k,i}) \leq \epsilon, \quad \forall\, k\in \mathcal{K},\; i\in \hat{\mathcal{Q}}_k,
\tag{\ref{eq:main}b}\label{eq:main_b}\\[1ex]
& \|\mathbf{W}_k\|_F^2 \leq P_{\mathrm{t}}, \quad \forall\, k\in \mathcal{K},
\tag{\ref{eq:main}c}\label{eq:main_c}\\[1ex]
& \sum_{j\in \mathcal{N}} u_{k,j} \geq 1, \quad \forall\, k\in \mathcal{K},
\tag{\ref{eq:main}d}\label{eq:main_d}\\[1ex]
& \sum_{k\in \mathcal{K}} u_{k,j} = 1, \quad \forall\, j\in \mathcal{N},
\tag{\ref{eq:main}e}\label{eq:main_e}\\[1ex]
& u_{k,j}\in\{0,1\}, \quad \forall\, k\in \mathcal{K},\; \forall\, j\in \mathcal{N},
\tag{\ref{eq:main}f}\label{eq:main_f}
\end{align}
\end{subequations}
where $\Gamma_{\rm{t}}$ in \eqref{eq:main_a} denotes the radar SNR threshold, $\epsilon$ in \eqref{eq:main_b} denotes the DoA estimation CRB threshold, and $P_{\rm{t}}$ denotes the maximum BS transmit power budget. Constraints \eqref{eq:main_a} and \eqref{eq:main_b} represent the sensing performance requirements. Constraint \eqref{eq:main_c} ensures that the BS transmit power does not exceed the maximum power budget. Constraint \eqref{eq:main_d} indicates that each BS should serve at least one ground CU. Constraint \eqref{eq:main_e} ensures that each CU can only be served by a single BS. 

Problem (P1) is a highly non-convex mixed-integer optimization problem including continuous variables $\{\mathbf{W}_k\}^K_{k=1}$ and integer variables $\mathbf{U}$ due to the highly coupled optimization problem \eqref{eq:main} and highly non-convex constraints \eqref{eq:main_a}, \eqref{eq:main_b}, and \eqref{eq:main_f}. Thus, it is generally challenging to directly obtain a global optimal solution. To tackle this challenging problem, we decompose the original problem (P1) into two sub-problems and alternatively solve them via AO.

\subsection{Problem Decomposition}
To efficiently solve problem (P1), we alternatively solve two sub-problems in two independent phases: the UA phase and the beamforming optimization phase. We optimize the two phases alternately in each outer iteration until convergence. First, given initial BS beamforming matrices $\{\mathbf{W}^{(0)}_k\}^K_{k=1}$, we obtain the initial UA strategy by solving sub-problem (P2), which can be written as
\begin{subequations}\label{eq:main_2}
% 主公式部分
\begin{align}
\text{(P2):}\quad \max_{\mathbf{U}} 
&\; \sum_{i\in \mathcal{N}} \sum_{k\in \mathcal{K}} u_{k,i}\,\frac{B}{\|\widetilde{\mathbf{u}}_k\|_0}\times \notag \\&\, \log_2\Biggl(1+\frac{|\mathbf{h}^H_{k,i}\mathbf{w}^{(0)}_{k,i}|^2}
{\sum_{n\neq i}^{N+M} |\mathbf{h}^H_{k,i}\mathbf{w}^{(0)}_{k,n}|^2+\sigma^2_i}\Biggr)
\tag{\ref{eq:main_2}}\label{eq:main_obj_2}
\end{align}
\vspace{-1em}
% 约束部分
\begin{align}
~\text{s.t.}\quad
& \sum_{j\in \mathcal{N}} u_{k,j} \geq 1, \quad \forall\, k\in \mathcal{K},
\tag{\ref{eq:main_2}d}\label{eq:main_d_2}\\[1ex]
& \sum_{k\in \mathcal{K}} u_{k,j} = 1, \quad \forall\, j\in \mathcal{N},
\tag{\ref{eq:main_2}e}\label{eq:main_e_2}\\[1ex]
& u_{k,j}\in\{0,1\}, \quad \forall\, k\in \mathcal{K},\; \forall\, j\in \mathcal{N}.
\tag{\ref{eq:main_2}f}\label{eq:main_f_2}
\end{align}
\end{subequations}

Given the obtained UA strategy $\mathbf{U}^*$, the BS transmit beamforming matrix optimization sub-problem can be written as
\begin{subequations}\label{eq:main_3}
% 主公式部分
\begin{align}
\text{(P3):}\quad \max_{\{\mathbf{W}_k\}_{k=1}^K} 
&\; \sum_{i\in \mathcal{N}} \sum_{k\in \mathcal{K}} u^*_{k,i}\,\frac{B}{\|\widetilde{\mathbf{u}}^*_k\|_0}\times \notag \\&\, \log_2\Biggl(1+\frac{|\mathbf{h}^H_{k,i}\mathbf{w}_{k,i}|^2}
{\sum_{n\neq i}^{N+M} |\mathbf{h}^H_{k,i}\mathbf{w}_{k,n}|^2+\sigma^2_i}\Biggr)
\tag{\ref{eq:main_3}}\label{eq:main_obj_3}
\end{align}
\vspace{-1em}
% 约束部分
\begin{align}
~\text{s.t.}\quad
& \frac{\sigma^2_t|\tilde{\beta}_{k,i}|^2\,\mathbf{W}^H_k\,\mathbf{g}_{k,i}\mathbf{g}^H_{k,i}\,\mathbf{W}_k}{\sigma^2_r}
\geq \Gamma_\text{t}, \nonumber ~ \forall\, k\in \mathcal{K},\; i\in \tilde{\mathcal{Q}}_k,
\tag{\ref{eq:main_3}a}\label{eq:main_a_3}\\[1ex]
& \operatorname{CRB}(\phi_{k,i}) \leq \epsilon, \quad \forall\, k\in \mathcal{K},\; i\in \hat{\mathcal{Q}}_k,
\tag{\ref{eq:main_3}b}\label{eq:main_b_3}\\[1ex]
& \|\mathbf{W}_k\|_F^2 \leq P_{\mathrm{t}}, \quad \forall\, k\in \mathcal{K}.
\tag{\ref{eq:main_3}c}\label{eq:main_c_3}
\end{align}
\end{subequations}

\textcolor{black}{To realize such a coordinated multi-cell optimization framework, we assume there exists a remote cloud server acting as a control server while the multiple BSs act as edge servers. The cloud controller and edge BSs form a hierarchical network. Specifically, the cloud server gathers comprehensive cross-cell information, including channel state information (CSI), user association patterns, and precoding/beamforming matrices from all edge BSs participating in coordinated beamforming. The cloud server analyzes the collected data, executes optimization algorithms, and distributes the optimized solutions back to the edge BSs for local execution. The edge BSs then implement the instructions obtained from the connected cloud server via an optical backhaul link in real-time, dynamically managing local beamforming and resource allocation decisions to ensure optimized performance in a rapidly changing wireless environment.}

\section{LLM-Enabled AO-Based Algorithm Design}
To solve problem (P1), we decompose it into two sub-problems (P2) and (P3) and apply an AO-based algorithm to iteratively optimize UA and BS transmit beamforming by combining LLMs and convex-based optimization as follows:

\subsection{LLM-based User Association Optimization}
In what follows, we propose an LLM-based framework to solve the UA strategy optimization sub-problem (P2) given the BS beamforming matrices. Problem (P2) is a binary integer programming problem that cannot be directly solved due to the huge number of possible UA combinations for large $N$ and $K$. To efficiently optimize $\mathbf{U}$, inspired by the substantial mathematical capabilities of LLMs,  we utilize LLMs as a black-box solver for problem (P2). 

\textcolor{black}{Fine-tuning Pre-trained Foundation Models (PFMs), including Large Language Models (LLMs), is expected to be instrumental in facilitating intelligent, hierarchical coordination across future 6G networks. These networks are envisioned to span cloud infrastructures, edge servers such as base stations or intelligent edge nodes, and a wide range of user equipment. To enable such coordination, the fine-tuning process should begin with the development of diversified datasets that capture the heterogeneity of 6G environments, including dynamic mobility patterns, variable quality-of-service (QoS) demands, and stringent resource constraints. The fine-tuning of PFMs should be approached in a hierarchical manner, where cloud infrastructures focus on global optimization tasks such as strategic resource management and interference coordination. In parallel, edge servers are responsible for context-specific functionalities, such as low-latency task execution and localized sensing. At the user level, models are expected to provide personalized services and adapt to the immediate environment of the user. } 
\textcolor{black}{Given the dynamic nature of 6G networks, techniques such as prompt engineering allow PFMs to generalize to new conditions with minimal retraining. To utilize LLMs resource-efficiently and flexibly for solving problem (P2), we aim to design prompt engineering to adapt LLMs for optimization using several techniques.}

\subsubsection{Motivation for Prompt Engineering}
The design of effective prompts is critical for guiding LLMs so that their outputs are relevant, consistent, and aligned with user intent. Various prompting methods embody distinct principles and are suitable for different scenarios. In this work, we aim to design a comprehensive prompt for solving UA strategy optimization sub-problem (P2). We first utilize in-context prompting along with expert knowledge to provide background knowledge for LLMs. Then, we utilize the zero-shot chain of thought techniques to ask LLMs to give an initial solution to the optimization problem. Following the idea of self-reflection, we ask LLMs, provided with few-shot prior examples, to self-enhance their solutions iteratively until the objective function value converges. By comprehensively and strategically applying these diverse prompting strategies, we can tailor prompts to meet the specific demands of a given task, thereby optimizing the performance and quality of LLM output. For illustration purposes, we briefly introduce the prompt techniques used in our design.

First, in-context prompting embeds pertinent background information directly into the prompt to ensure that LLM responses are enriched with relevant context, which is useful for tasks requiring in-depth analysis. Moreover, expert knowledge ensures that the embedded background information is both accurate and relevant to the domain at hand, which allows the prompts to precisely guide the LLMs \cite{WirelessLLM}. 

Second, zero-shot prompting uses clear and direct instructions without providing examples, which is suitable for simple tasks or situations that demand prompt responses. However, it may encounter difficulties when addressing ambiguous requirements in more complex cases \cite{LLM1}. In contrast, few-shot prompting improves clarity by incorporating exemplar inputs that define the desired output format, which is suitable for tasks with nuanced requirements.

Third, chain of thought prompting encourages LLMs to articulate intermediate reasoning steps that lead to the final answer to one complex problem. The chain of thought provides an interpretable window for LLMs to reach the correct answer and debug where the reasoning path might go wrong \cite{LLM2}.

Finally, the self-reflection technique has proven to be a powerful tool that enables LLMs to learn from their mistakes, self-correct, and generate more accurate outputs \cite{LLM3}. For complex problem-solving, self-reflection is especially beneficial when dealing with tasks that require multi-step reasoning or synthesizing information from multiple sources. By evaluating their initial approaches and identifying errors, LLMs can refine their strategies and produce more precise solutions, ultimately leading to improved results.

\subsubsection{Prompting-Based LLM as a Black-Box Optimizer}
\begin{figure*}[p] 
\centering
 \makebox[\textwidth]{\includegraphics[width=.8\paperwidth]{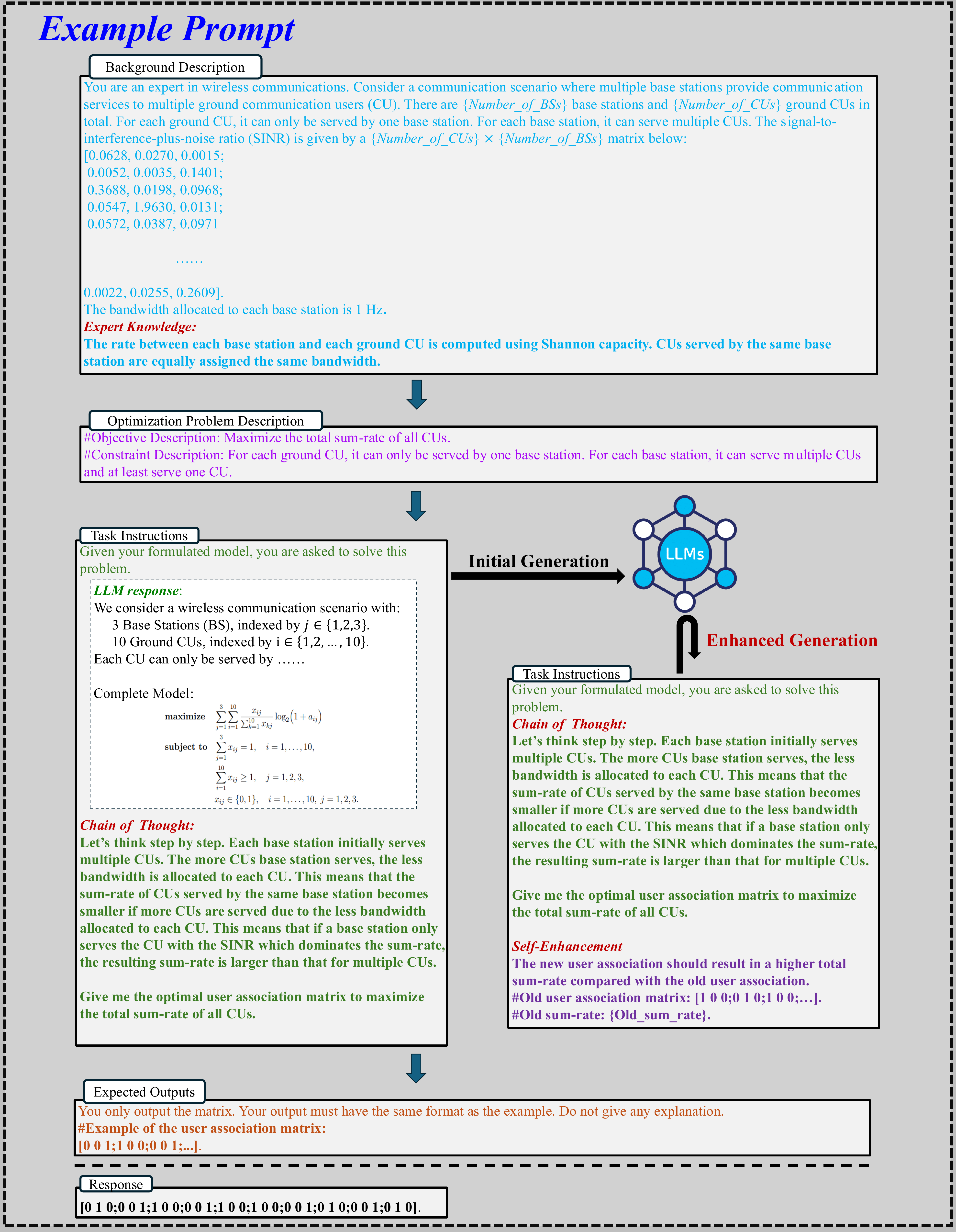}}
\caption{An example prompt for solving sub-problem (P2) with $K=3$ BSs and $N=10$ ground CUs.}
\label{prompt}
\end{figure*}

We aim to design prompt engineering using the techniques of in-context learning and chain of thought in an iterative self-reflection manner, which enables LLMs to generate a mathematical model of our optimization problem (P2) based on the prompt and find a near-optimal UA strategy. The structure of the designed prompting involves background description, optimization problem description, task instructions, self-enhancement, and expected outputs, as follows (also shown in Fig. \ref{prompt}).

a) Background description:
In this part, we aim to provide LLMs with the necessary background knowledge of our problem. Specifically, as shown in Fig. \ref{prompt}, we first specify the role of LLMs as experts in wireless communications. Consequently, LLMs are expected to understand the following question using their expert knowledge of wireless communications. Then, we introduce the system settings of our problem to the LLMs, including the number of BSs and ground CUs and the UA rules. Such information provides LLMs with a clear picture of our system, which helps them better understand the optimization problem. Moreover, to facilitate LLMs to accurately generate the mathematical model of the optimization problem (P2), we provide LLMs with extra expert knowledge as highlighted in Fig. \ref{prompt}. Specifically, we instruct the LLMs to calculate the communication rate from each BS to each CU using the Shannon capacity while considering the bandwidth allocation of multiple CUs. This background context serves as a foundation for guiding LLMs to better understand the problem and reduce ambiguity in the instructions, which can effectively set the stage for in-context learning.

b) Optimization problem description:
In this part, given the specified background descriptions and expert knowledge, we elaborate optimization problem (P2) with natural language for LLMs. To help LLMs better distinguish between the objective function and constraints, we highlight the objective and constraint descriptions using special marks, which can clarify and enhance the structure of the prompt.

c) Task instructions:
In this part, we clarify the task instructions for LLMs. First, we instruct the LLMs to formulate the mathematical model of our optimization problem (P2). Given the background knowledge and a clearly structured problem description, LLMs can accurately formulate problem (P2), as shown in Fig. \ref{prompt}. Then, we instruct LLMs to solve the formulated problem to generate a near-optimal UA strategy. To achieve this goal, we utilize the chain of thought technique to guide the LLMs, as highlighted in Fig. \ref{prompt}. More specifically, the chain of thought guides LLMs in solving the optimization problem step by step. LLMs are asked to consider that the sum rate of the CUs served by the same base station becomes smaller if more CUs are served. Then, the LLMs consider that the BS should give priority to the CU with the largest SINR if the sum rate is dominated by such a CU to maximize the total sum rate. Then, the LLMs are supposed to find the optimal UA strategy to maximize the objective function under the above guidance. The chain of thought prompting encourages LLMs to break down the reasoning process into intermediate steps, which can lead to more accurate and robust problem solving.

d) Self-enhancement:
Initially, the LLMs may generate unexpected or incorrect solutions. To tackle this issue, in this part, we aim to make LLMs iteratively self-refine the solutions obtained in previous iterations as highlighted in Fig. \ref{prompt}. Specifically, in each iteration, we provide the prior UA strategy and the objective function value from the previous iteration to guide LLMs to self-refine their solutions until the objective function value exceeds that from the last iteration or the objective function value meets the convergence criteria.

e) Expected outputs:
In this part, we describe the expected output solutions. The output of LLM should only be given in our expected format. As shown in Fig. \ref{prompt}, the output response from LLMs is given exactly in the expected format. 
\subsubsection{Algorithm Summary}
Given the above designs, the proposed framework of the prompting-based black-box optimizer is summarized in Algorithm \ref{algorithm1}.
\begin{algorithm}[htb]
    \caption{Prompt Engineering Structure for UA Optimization in Problem (P2)}
    \label{algorithm1}
    \KwIn{Number of BSs: \{\textit{Number\_of\_BSs}\}. Number of CUs: \{\textit{Number\_of\_CUs}\}. SINR matrix: \{\textit{SINR\_Matrix}\}.}
    \KwOut{UA strategy $\mathbf{U}$ }
    \textbf{Step 1: Background Description}:\\
    Provide the background description of the problem, together with expertise.\\
    \textbf{Step 2: Optimization Problem Description}:\\
    Provide the optimization problem description based on Problem (P2).\\
    \textbf{Step 3: Task Instructions}:\\
    Give task instructions to obtain the initial solution.\\
    \While{no convergence}{Self-enhance the task instructions to obtain the refined solution.}
    \textbf{Step 4: Expected Outputs}:\\
    LLMs generate outputs in the expected format.\\
    \Return{Final UA strategy $\mathbf{U}^*$}.
    \end{algorithm}
    
\subsection{ADMM-Based BS Transmit Beamforming Optimization}
In what follows, we propose an ADMM-based algorithm to solve the BS transmit beamforming optimization sub-problem (P3) given a fixed UA strategy $\mathbf{U}^*$. It is clear that problem (P3) is highly non-convex even with fixed integer variables due to the $\rm{log}_2(\cdot)$ in the objective function \eqref{eq:main_3} and constraints \eqref{eq:main_a_3}, \eqref{eq:main_b_3}. To effectively optimize $\mathbf{W}_k,~k=1,\dots,K$, for each BS $k$, we further decompose the problem (P3) into a series of sub-problems and solve them simultaneously, where each sub-problem omitting bandwidth $B$ is written as
\begin{subequations}\label{eq:main_4}
% 主公式部分
\begin{align}
\text{(P3.$k$):}\quad &\max_{\mathbf{W}_k} 
\; \sum_{i\in \mathcal{N}} u^*_{k,i}\,\frac{1}{\|\widetilde{\mathbf{u}}^*_k\|_0}\times \notag \\&\, \log_2\Biggl(1+\frac{|\mathbf{h}^H_{k,i}\mathbf{w}_{k,i}|^2}
{\sum_{n\neq i}^{N+M} |\mathbf{h}^H_{k,i}\mathbf{w}_{k,n}|^2+\sigma^2_i}\Biggr)
\tag{\ref{eq:main_4}}\label{eq:main_obj_4}
\end{align}
\vspace{-1em}
% 约束部分
\begin{align}
~\text{s.t.}\quad
& \frac{\sigma^2_t|\tilde{\beta}_{k,i}|^2\,\mathbf{W}^H_k\,\mathbf{g}_{k,i}\mathbf{g}^H_{k,i}\,\mathbf{W}_k}{\sigma^2_r}
\geq \Gamma_\text{t}, \quad i\in \tilde{\mathcal{Q}}_k,
\tag{\ref{eq:main_4}a}\label{eq:main_a_4}\\[1ex]
& \operatorname{CRB}(\phi_{k,i}) \leq \epsilon, \quad i\in \hat{\mathcal{Q}}_k,
\tag{\ref{eq:main_4}b}\label{eq:main_b_4}\\[1ex]
& \|\mathbf{W}_k\|_F^2 \leq P_{\mathrm{t}}.
\tag{\ref{eq:main_4}c}\label{eq:main_c_4}
\end{align}
\end{subequations}
To solve this problem, we propose an algorithm framework based on FP, MM, and ADMM to decompose problem (P3.$k$) into a series of sub-problems and solve them iteratively until convergence.
\subsubsection{Problem reformulation based on FP}
Based on \cite{FP1}, we introduce a Lagrangian dual reformulation of problem (P3.$k$) by introducing a set of auxiliary variables $\boldsymbol{\upsilon}=[\upsilon_1, \upsilon_2, \cdots, \upsilon_N]^T$. The reformulated objective function is given as follows
\begin{align}
    {\Phi}(\mathbf{W}_k, \boldsymbol{\upsilon})=\sum^N_{i=1}&\omega_i\text{log}_2(1+\upsilon_i)-\sum^N_{i=1}\omega_i\upsilon_i\notag \\&+\sum^N_{i=1}\frac{\omega_i(1+\upsilon_i)|\mathbf{h}^H_{k,i}\mathbf{w}_{k,i}|^2}{\sum^{N+M}_{n=1}|\mathbf{h}^H_{k,i}\mathbf{w}_{k,n}|+\sigma^2_i},
\end{align}
where $\omega_i\triangleq u^*_{k,i}\,\frac{1}{\|\widetilde{\mathbf{u}}^*_k\|_0}$, and variable $\mathbf{W}_k$ is taken out of $\rm{log}_2(\cdot)$ and included in the third term. A detailed proof of the problem reformulation is provided in Part \Rmnum{2} of \cite{FP2}. Given a fixed $\boldsymbol{\upsilon}$, to optimize $\mathbf{W}_k$, we expand the third quadratic form and further recast ${\Phi}(\mathbf{W}_k, \boldsymbol{\upsilon})$ as 
\begin{align}
    {\Phi}(\mathbf{W}_k, \boldsymbol{\upsilon}, \boldsymbol{b})=&\sum^N_{i=1}\omega_i\text{log}_2(1+\upsilon_i)-\sum^N_{i=1}\omega_i\upsilon_i\notag \\ &+\sum^N_{i=1}2\text{Re}\{\bar{b}_i\mathbf{h}^H_{k,i}\mathbf{w}_{k,i}\}\sqrt{\omega_i(1+\upsilon_i)}\notag\\
    &-\sum^N_{i=1}|b_i|^2\left(\sum^{N+M}_{n=1}|\mathbf{h}^H_{k,n}\mathbf{w}_{k,n}|^2+\sigma^2_i\right),
\end{align}
where $\boldsymbol{b}=[b_1,b_2,\cdots,b_N]^T$ denotes introduced auxiliary variables. To simplify the reformulated objective function ${\Phi}(\mathbf{W}_k, \boldsymbol{\upsilon}, \boldsymbol{b})$, we introduce vector variable $\boldsymbol{w}_k\in \mathbb{C}^{M(M+N)\times1}$ by vertically stacking the $\mathbf{w}_{k,i}, i=1,2,\cdots,N+M$. Therefore, objective function ${\Phi}(\mathbf{W}_k, \boldsymbol{\upsilon}, \boldsymbol{b})$ can be equivalently expressed as
\begin{equation}
    {\Phi}(\boldsymbol{w}_k, \boldsymbol{\upsilon}, \boldsymbol{b})=\text{Re}\{\mathbf{f}^H\boldsymbol{w}_k\}-||\mathbf{F}\boldsymbol{w}_k||_2^2+\delta(\boldsymbol{\upsilon}, \boldsymbol{b}), \label{eq:33}
\end{equation}
where $\delta(\boldsymbol{\upsilon},\boldsymbol{b})$ collects all forms that do not depend on $\boldsymbol{\upsilon}$ and $\boldsymbol{b}$. Vector $\mathbf{f}$ is given as follows 
\begin{equation}
\begin{aligned}
&\mathbf{f} 
\;\triangleq
\notag \\ &\Bigl[
2\sqrt{\omega_1\bigl(1+\upsilon_1\bigr)}\,\bar{b}_1\mathbf{h}_{k,1}^H,\;\dots,\;2\sqrt{\omega_N\bigl(1+\upsilon_N\bigr)}\,\bar{b}_N\mathbf{h}_{k,N}^H,\notag \\&
\mathbf{0}^H,\;\dots,\;\mathbf{0}^H
\Bigr]^H
\;\in\;
\mathbb{C}^{\,M(N+M)\times1},
\end{aligned}
\end{equation}
where $\mathbf{0}\triangleq[0,0,\cdots,0] \in \mathbb{C}^{M\times1}$ denotes the vector of all zeros. Matrix $\mathbf{F}$ in \eqref{eq:33} is given by
\begin{equation}
\mathbf{F}=[\mathbf{F}^T_1, \mathbf{F}^T_2, \cdots, \mathbf{F}^T_N]^T
\quad\in\;\mathbb{C}^{\,N(N+M)\,\times\,M(N+M)},
\end{equation}
where $\mathbf{F}_i= \mathbf{I}_{N+M}\otimes\bar{b}_{i}\,\mathbf{h}_{k,i}^{H}, i=1,\cdots,N.$
% \\
% \quad F_i &= \begin{bmatrix}
% \bar{b}_{i}\,\mathbf{h}_{k,i}^{H} & \mathbf{0} & \cdots & \mathbf{0}\\[3pt]
% \mathbf{0} & \bar{b}_{i}\,\mathbf{h}_{k,i}^{H} & \cdots & \mathbf{0}\\[2pt]
% \vdots & \vdots & \ddots & \vdots\\[2pt]
% \mathbf{0} & \mathbf{0} & \cdots & \bar{b}_{i}\,\mathbf{h}_{k,i}^{H}
% \end{bmatrix}\\
% &\quad\in\;\mathbb{C}^{(N+M)\times\,M(N+M)},
% \quad
% i=1,\dots,N.
\subsubsection{Constraint transformation}
The radar SNR constraint \eqref{eq:main_a_4} can be rewritten as
\begin{equation}
    \sum^{N+M}_{j=1}\mathbf{w}^H_{k,j}\mathbf{g}_{k,i}\mathbf{g}^H_{k,i}\mathbf{w}_{k,j} \geq \tilde{\Gamma}_t, \quad i\in\tilde{\mathcal{Q}}_k,
\end{equation}
where $\tilde{\Gamma}_t=\frac{\Gamma_t\sigma^2_r}{\sigma^2_t|\tilde{\beta}_{k,i}|^2}$. It can be equivalently transformed with respect to $\boldsymbol{w}_k$ as
\begin{equation}
    \boldsymbol{w}_k^H\mathcal{G}_i^H\mathcal{G}_i\boldsymbol{w}_k \geq \tilde{\Gamma}_t, i \in \tilde{\mathcal{Q}}_k, \label{SNR2}
\end{equation}
where $\mathcal{G}_i=\mathbf{I}_{N+M}\otimes\mathbf{g}^H_{k,i}$.

Noting that constraint \eqref{SNR2} is not convex, we apply the first-order Taylor expansion with respect to $\boldsymbol{w}_k$ on \eqref{SNR2} and obtain 
\begin{align}
    \boldsymbol{w}_k^H\mathcal{G}_i^H\mathcal{G}_i\boldsymbol{w}_k \geq  &-(\boldsymbol{w}_k^{(j)})^H\mathcal{G}_i^H\mathcal{G}_i\boldsymbol{w}_k^{(j)}\notag\\&+2\text{Re}\{(\boldsymbol{w}_k^{(j)})^H\mathcal{G}_i^H\mathcal{G}_i\boldsymbol{w}_k\},
\end{align}
where $\boldsymbol{w}_k^{(j)}$ denotes the acquired solution in the $j$-th iteration. Thus, 
the original constraint \eqref{eq:main_a_4} is transformed into 
\begin{equation}   -(\boldsymbol{w}_k^{(j)})^H\mathcal{G}_i^H\mathcal{G}_i\boldsymbol{w}_k^{(j)}+2\text{Re}\{(\boldsymbol{w}_k^{(j)})^H\mathcal{G}_i^H\mathcal{G}_i\boldsymbol{w}_k\} \geq \tilde{\Gamma_t} \label{SNR3}.
\end{equation}

To handle CRB constraint \eqref{eq:main_b_4}, we transform it into the following constraints based on \cite{wang}
\begin{align}
    &C^{-1} \leq \epsilon,\\
    &\left[\begin{array}{ll}
{J}(\boldsymbol{\xi}_{k,i})_{\phi_{k,i}\phi_{k,i}}-C & \mathbf{J}(\boldsymbol{\xi}_{k,i})_{\phi_{k,i}\boldsymbol{\alpha}^T_{i}} \\
\mathbf{J}(\boldsymbol{\xi}_{k,i})^T_{\phi_{k,i}\boldsymbol{\alpha}^T_{i}} & \mathbf{J}(\boldsymbol{\xi}_{k,i})_{\boldsymbol{\alpha}_{i}\boldsymbol{\alpha}^T_{i}}
\end{array}\right] \succeq 0, \quad i\in \hat{\mathcal{Q}}_k \label{crb1} \\
& C > 0,
\end{align}
where ${C} \in \mathbb{R}^+$ is an introduced auxiliary variable. However, constraint \eqref{crb1} is still non-convex and very challenging to handle. To address this challenge, we introduce an auxiliary variable $\mathbf{q}\triangleq[q_1,q_2,q_3]^T \in \mathbb{C}^{3\times1}$ to take $\mathbf{W}_k$ out of the positive semidefinite matrix (PSD) constraint \eqref{crb1}, where $q_i, ~i=1,2,3$ is expressed as
\begin{equation}
    q_i \triangleq Q_i(\mathbf{W}_k), \quad i=1,2,3,
\end{equation}
where $Q_i(\cdot)$ denotes the function with respect to $\mathbf{W}_k$ based on the FIM defined in \eqref{FIM1} and \eqref{FIM2}. The detailed definitions of $Q_i(\cdot), ~i=1,2,3$, are given in Appendix \ref{appenA}.

\subsubsection{ADMM-based augmented Lagrangian problem solution}
Based on the above derivations, problem (P3.$k$) is transformed into
\begin{subequations}\label{eq:main_5}
\begin{align}
& \max_{\boldsymbol{w}_k, \boldsymbol{\upsilon}, \boldsymbol{b}, C, \mathbf{q}}~{\Phi}(\boldsymbol{w}_k, \boldsymbol{\upsilon}, \boldsymbol{b})\\
\text{s.t.} &\quad-(\boldsymbol{w}_k^{(j)})^H\mathcal{G}_i^H\mathcal{G}_i\boldsymbol{w}_k^{(j)}\notag \\&+2\text{Re}\{(\boldsymbol{w}_k^{(j)})^H\mathcal{G}_i^H\mathcal{G}_i\boldsymbol{w}_k\} \geq \tilde{\Gamma}_t,~i \in \tilde{\mathcal{Q}}_k, \label{snr3}\\
    &C^{-1} \leq \epsilon,\\
    &\left[\begin{array}{ll}
q_1-C & \text{Re}\{q_2[1~\mathfrak{j}]\} \\
\text{Re}\left\{q_2\left[\begin{array}{l}
1 \\
\mathfrak{j}
\end{array}\right]\right\} & q_3\mathbf{I}_2
\end{array}\right] \succeq 0, ~i\in \hat{\mathcal{Q}}_k, \notag \\
& C > 0, \\
& ||\boldsymbol{w}_k||^2_2 \leq P_t, \label{pt1}\\
& q_i = Q_i(\mathbf{W}_k),~ i=1,2,3. \label{eq_cons1}
\end{align}
\end{subequations}
To tackle constraints \eqref{eq_cons1}, we formulate the augmented Lagrangian problem of \eqref{eq:main_5} by introducing the Lagrangian dual variable $\mathbf{\Upsilon}\triangleq[\Upsilon_1, \Upsilon_2, \Upsilon_3]^T \in \mathbb{C}^{3\times1}$ and the penalty factor $\rho>0$ as \cite{AL}
\begin{align}
\mathcal{L}(\rho, \mathbf{\Upsilon}): &\min_{\boldsymbol{w}_k, \boldsymbol{\upsilon}, \boldsymbol{b}, C, \mathbf{q}} -{\Phi}(\boldsymbol{w}_k, \boldsymbol{\upsilon}, \boldsymbol{b}) + \mathcal{P}_{\rho}(\mathbf{W}_k, \mathbf{\Upsilon})\\
& \text { s.t. }\, \eqref{snr3}-\eqref{pt1},
\end{align}
where 
\begin{equation}
    \mathcal{P}_{\rho}(\mathbf{W}_k, \mathbf{\Upsilon})=\frac{1}{2\rho}\sum^3_{i=1}|Q_i(\mathbf{W}_k)-q_i+\rho\Upsilon_i |^2.
\end{equation}
To solve the augmented Lagrangian problem, we first optimize the original problem using block coordinate descent (BCD) in the inner loop and then update the Lagrangian dual variables and the penalty factor in the outer loop, as presented below.

a) Update $\boldsymbol{\upsilon}$ and $\boldsymbol{b}$:
For given $\mathbf{W}_k$, ${\Phi}(\mathbf{W}_k,\boldsymbol{\upsilon})$ is convex with respect to $\boldsymbol{\upsilon}$. We can obtain the optimal $\boldsymbol{\upsilon}^*$ by setting $\partial{\Phi}(\boldsymbol{\upsilon})/\partial\boldsymbol{\upsilon}$ to zero. Thus, we have
\begin{equation}
    \upsilon^*_i = \frac{|\mathbf{h}^H_{k,i}\mathbf{w}_{k,i}|^2}{\sum^{N+M}_{n=1, n\neq i}|\mathbf{h}^H_{k,i}\mathbf{w}_{k,n}|+\sigma^2_i}, ~\forall i \in \mathcal{N}. \label{update_v}
\end{equation}
Note that $\upsilon^*_i$ is equal to the received SINR from BS $k$ at CU $i$. 
Similarly, to obtain the optimal $\boldsymbol{b}^*$ for given $\mathbf{W}_k$ and optimal $\boldsymbol{\upsilon}^*$, we set $\partial{\Phi}(\mathbf{W}_k, \boldsymbol{\upsilon}, \boldsymbol{b})/\partial\boldsymbol{b}$ equal to zero, and obtain
\begin{equation}
    b^*_i=\frac{\sqrt{\omega_i(1+\upsilon^*_i)}\mathbf{h}^H_{k,i}\mathbf{w}_{k,i}}{\sum^{N+M}_{n=1}|\mathbf{h}^H_{k,i}\mathbf{w}_{k,n}|+\sigma^2_i}, ~\forall i \in \mathcal{N}. \label{update_b}
\end{equation}

b) Sub-problem with respect to $C$ and $\mathbf{q}$:
For given $\boldsymbol{w}_k, \boldsymbol{\upsilon},$ and $\boldsymbol{b}$, the optimization of $C$ and $\mathbf{q}$ can be expressed as the following sub-problem
\begin{subequations} \label{sub_Cq}
\begin{align} 
    \min_{C, \mathbf{q}}\quad&\frac{1}{2\rho}\sum^3_{i=1}|Q_i(\mathbf{W}_k)-q_i+\rho\Upsilon_i |^2 \\
    \text{s.t.} \quad &C^{-1} \leq \epsilon,\\
    &\left[\begin{array}{ll}
q_1-C & \text{Re}\{q_2[1~\mathfrak{j}]\} \\
\text{Re}\left\{q_2\left[\begin{array}{l}
1 \\
\mathfrak{j}
\end{array}\right]\right\} & q_3\mathbf{I}_2
\end{array}\right] \succeq 0, ~i\in \hat{\mathcal{Q}}_k, \notag \\
& C \succ 0.
\end{align}
\end{subequations}
Note that problem \eqref{sub_Cq} is a semidefinite problem (SDP) and can be solved optimally using existing optimization toolboxes such as CVX \cite{cvx}.

c) Sub-problem with respect to $\boldsymbol{w}_k$:
The sub-problem with respect to $\boldsymbol{w}_k$ is given as
\begin{subequations}
    \begin{align} 
    \min_{\boldsymbol{w}_k}\quad &||\mathbf{F}\boldsymbol{w}_k||_2^2-\text{Re}\{\mathbf{f}^H\boldsymbol{w}_k\} \notag\\&+ \frac{1}{2\rho}\sum^3_{i=1}|\text{Tr}\{\boldsymbol{\Theta}_i\mathbf{W}_k\mathbf{W}^H_k\}+\mathfrak{c}_i|^2 \label{sub_w}\\
    \text{s.t.}\quad &-(\boldsymbol{w}_k^{(j)})^H\mathcal{G}_i^H\mathcal{G}_i\boldsymbol{w}_k^{(j)}\notag \\&+2\text{Re}\{(\boldsymbol{w}_k^{(j)})^H\mathcal{G}_i^H\mathcal{G}_i\boldsymbol{w}_k\} \geq \tilde{\Gamma}_t,~i \in \tilde{\mathcal{Q}}_k,\\
& ||\boldsymbol{w}_k||_2^2 \leq P_t,
    \end{align}
\end{subequations}
where the definitions of $\boldsymbol{\Theta}_i, ~i=1,2,3$, are given in Appendix \ref{appenA}, and we introduce constants $\mathfrak{c}_i\triangleq-q_i+\rho\Upsilon_i, ~i=1,2,3$. The sub-problem is non-convex due to the third term in \eqref{sub_w}. To address this challenge, we apply the MM method to construct surrogate functions that locally approximate the objective function and then minimize the surrogate functions.

To start with, we first expand the non-convex term as $|\text{Tr}\{\boldsymbol{\Theta}_i\mathbf{W}_k\mathbf{W}^H_k\}+\mathfrak{c}_i|^2, ~i=1,2,3$, which can be expressed as
\begin{align}
    &\quad~|\text{Tr}\{\boldsymbol{\Theta}_i\mathbf{W}_k\mathbf{W}^H_k\}+\mathfrak{c}_i|^2 \notag \\
    &=2\text{Re}\left\{\bar{\mathfrak{c}}_i\text{Tr}\{\boldsymbol{\Theta}_i\mathbf{W}_k\mathbf{W}^H_k\}\right\}+|\text{Tr}\{\boldsymbol{\Theta}_i\mathbf{W}_k\mathbf{W}^H_k\}|^2+|\mathfrak{c}_i|^2 \notag \\
&=2\text{Re}\left\{\bar{\mathfrak{c}}_i\boldsymbol{w}_k^H(\mathbf{I}_{N+M}\otimes\boldsymbol{\Theta}_i)\boldsymbol{w}_k\right\}
\notag \\&+|\boldsymbol{w}_k^H(\mathbf{I}_{N+M}\otimes\boldsymbol{\Theta}_i)\boldsymbol{w}_k|^2+|\mathfrak{c}_i|^2,~ i=1,2,3,\label{surr0}
\end{align}
where the expansion is based on the transformation $\text{Tr}\{\mathbf{A B C D}\}=\text{vec}^H\left\{\mathbf{D}^H\right\}\left(\mathbf{C}^T \otimes \mathbf{A}\right) \text{vec}\{\mathbf{B}\}$, and $\otimes$ denotes the Kronecker product operation. We rewrite the first term in \eqref{surr0} as follows
\begin{align}
    &\quad ~2\text{Re}\left\{\bar{\mathfrak{c}}_i\boldsymbol{w}_k^H(\mathbf{I}_{N+M}\otimes\boldsymbol{\Theta}_i)\boldsymbol{w}_k\right\}\notag \\
    &=\bar{\mathfrak{c}}_i\boldsymbol{w}_k^H(\mathbf{I}_{N+M}\otimes\boldsymbol{\Theta}_i)\boldsymbol{w}_k+\mathfrak{c}_i\boldsymbol{w}_k^H(\mathbf{I}_{N+M}\otimes\boldsymbol{\Theta}^H_i)\boldsymbol{w}_k \notag \\
    &=\boldsymbol{w}_k^H\tilde{\boldsymbol{\Theta}}_i\boldsymbol{w}_k,
\end{align}
where $\tilde{\boldsymbol{\Theta}}_i\triangleq\bar{\mathfrak{c}}_i\times\mathbf{I}_{N+M}\otimes\boldsymbol{\Theta}_i+\mathfrak{c}_i\times\mathbf{I}_{N+M}\otimes\boldsymbol{\Theta}^H_i$ is a Hermitian matrix based on the properties of the Kronecker product. Therefore, the quadratic form $\boldsymbol{w}_k^H\tilde{\boldsymbol{\Theta}}_i\boldsymbol{w}_k$ can be upper-bounded (e.g., similar to Example 13 in \cite{MM1}) as
\begin{align}
    \boldsymbol{w}_k^H\tilde{\boldsymbol{\Theta}}_i\boldsymbol{w}_k &\leq \boldsymbol{w}_k^H\mathbf{V}\boldsymbol{w}_k+ 2\text{Re}\left\{\boldsymbol{w}_k^H(\tilde{\boldsymbol{\Theta}}_i-\mathbf{V})\boldsymbol{w}_k^{(j)}\right\}\notag\\ &+(\boldsymbol{w}_k^{(j)})^H(\mathbf{V}-\tilde{\boldsymbol{\Theta}}_i)\boldsymbol{w}_k^{(j)}, \label{surr1}
\end{align}
where $\mathbf{V}\triangleq\lambda_i\mathbf{I}_{M(N+M)} \succeq \tilde{\boldsymbol{\Theta}}_i$ with $\lambda_i$ being the maximum eigenvalue of $\tilde{\boldsymbol{\Theta}}_i$, and $\boldsymbol{w}_k^{(j)}$ denotes the acquired solution in the $j$-th iteration. Since $||\boldsymbol{w}_k||^2_2\leq P_t$, \eqref{surr1} is transformed into
\begin{align}
    \boldsymbol{w}_k^H\tilde{\boldsymbol{\Theta}}_i\boldsymbol{w}_k\leq &2\text{Re}\left\{\boldsymbol{w}_k^H(\tilde{\boldsymbol{\Theta}}_i-\mathbf{V})\boldsymbol{w}_k^{(j)}\right\}+ \lambda_iP_t \notag \\&+(\boldsymbol{w}_k^{(j)})^H(\mathbf{V}-\tilde{\boldsymbol{\Theta}}_i)\boldsymbol{w}_k^{(j)},
\end{align}
where the last two terms are constant with respect to $\boldsymbol{w}_k$. 

As for the third term in \eqref{surr0}, we expand it as
\begin{align}
    & \quad ~|\boldsymbol{w}_k^H(\mathbf{I}_{N+M}\otimes\boldsymbol{\Theta}_i)\boldsymbol{w}_k|^2 \notag \\
    &=\text{Tr}\left\{(\mathbf{I}_{N+M}\otimes\boldsymbol{\Theta}_i)\boldsymbol{w}_k\boldsymbol{w}_k^H(\mathbf{I}_{N+M}\otimes\boldsymbol{\Theta}_i)^H\boldsymbol{w}_k\boldsymbol{w}_k^H\right\} \notag \\
    &=\text{vec}^H\{\boldsymbol{w}_k\boldsymbol{w}_k^H\}\notag \\&\times \left({(\overline{\mathbf{I}_{N+M}\otimes\boldsymbol{\Theta}_i})}\otimes(\mathbf{I}_{N+M}\otimes\boldsymbol{\Theta}_i)\right)\text{vec}\{\boldsymbol{w}_k\boldsymbol{w}_k^H\} \notag \\
    &=\hat{\boldsymbol{w}_k}^H\hat{\boldsymbol{\Theta}}_i\hat{\boldsymbol{w}_k},
\end{align}
where $\hat{\boldsymbol{w}_k}\triangleq\text{vec}\{\boldsymbol{w}_k\boldsymbol{w}_k^H\}$, and $\hat{\boldsymbol{\Theta}}_i\triangleq(\overline{\mathbf{I}_{N+M}\otimes\boldsymbol{\Theta}_i}\otimes(\mathbf{I}_{N+M}\otimes\boldsymbol{\Theta}_i)$. Based on Lemma 12 in \cite{MM1}, we have the following inequality
\begin{align}
    &\quad~~\hat{\boldsymbol{w}_k}^H\hat{\boldsymbol{\Theta}}_i\hat{\boldsymbol{w}_k} \notag\\
    &\leq \hat{\boldsymbol{w}_k}^H\hat{\mathbf{V}}\hat{\boldsymbol{w}_k}+\text{Re}\left\{(\hat{\boldsymbol{w}_k}^{(j)})^H(\hat{\boldsymbol{\Theta}}_i+\hat{\boldsymbol{\Theta}}_i^H-2\hat{\mathbf{V}})\hat{\boldsymbol{w}_k}\right\} \notag \\ &+(\hat{\boldsymbol{w}_k}^{(j)})^H(\hat{\mathbf{V}}-\hat{\boldsymbol{\Theta}}_i)\hat{\boldsymbol{w}_k}^{(j)}\notag \\
    & \leq \text{Re}\left\{(\hat{\boldsymbol{w}_k}^{(j)})^H(\hat{\boldsymbol{\Theta}}_i+\hat{\boldsymbol{\Theta}}_i^H-2\hat{\mathbf{V}}) \hat{\boldsymbol{w}_k}\right\} \notag \\&+\hat{\lambda}_iP^2_t+\hat{\lambda}_i(\hat{\boldsymbol{w}_k}^{(j)})^H\hat{\boldsymbol{w}_k}^{(j)}-(\hat{\boldsymbol{w}_k}^{(j)})^H\hat{\boldsymbol{\Theta}}_i\hat{\boldsymbol{w}_k}^{(j)}, \label{surr2}
\end{align}
where $\hat{\mathbf{V}}_i\triangleq\hat{\lambda}_i\mathbf{I}_{M^2(N+M)^2} \succeq \hat{\boldsymbol{\Theta}}_i$, and $\hat{\lambda}_i$ is the maximum eigenvalue of $\hat{\boldsymbol{\Theta}}_i$. We rewrite the first term in \eqref{surr2} and transform it as
\begin{align}
    &\quad~\text{Re}\left\{(\hat{\boldsymbol{w}_k}^{(j)})^H(\hat{\boldsymbol{\Theta}}_i+\hat{\boldsymbol{\Theta}}_i^H-2\hat{\mathbf{V}}) \hat{\boldsymbol{w}_k}\right\}\notag \\
    &=(\hat{\boldsymbol{w}_k}^{(j)})^H\hat{\boldsymbol{\Theta}}_i\hat{\boldsymbol{w}_k}+(\hat{\boldsymbol{w}_k}^{(j)})^H\hat{\boldsymbol{\Theta}}_i^H\hat{\boldsymbol{w}_k}-2(\hat{\boldsymbol{w}_k}^{(j)})^H\hat{\mathbf{V}}\hat{\boldsymbol{w}_k}, \notag \\
    &=|\boldsymbol{w}_k^H\text{vec}(\boldsymbol{\Theta}^H_i\mathbf{W}_k^{(j)})|^2+|\boldsymbol{w}_k^H\text{vec}(\boldsymbol{\Theta}_i\mathbf{W}_k^{(j)})|^2 \notag\\ &-2\hat{\lambda}_i\boldsymbol{w}_k^H\boldsymbol{w}_k^{(j)}(\boldsymbol{w}_k^{(j)})^H\boldsymbol{w}_k+\hat{\lambda}_iP^2_t+\hat{\lambda}_i(\hat{\boldsymbol{w}_k}^{(j)})^H\hat{\boldsymbol{w}_k}^{(j)} \notag \\&-(\hat{\boldsymbol{w}_k}^{(j)})^H\hat{\boldsymbol{\Theta}}_i\hat{\boldsymbol{w}_k}^{(j)}, \label{surr3}
\end{align}
where the first and second terms are convex, and the third term is concave. By applying the first-order Taylor expansion, we have
\begin{align}
    &\quad-2\hat{\lambda}_i\boldsymbol{w}_k^H\boldsymbol{w}_k^{(j)}(\boldsymbol{w}_k^{(j)})^H\boldsymbol{w}_k \notag \\
    &\leq -2\hat{\lambda}_i(||\boldsymbol{w}_k^{(j)}||^4 +2\text{Re}\{||\boldsymbol{w}_k^{(j)}||^2(\boldsymbol{w}_k^{(j)})^H(\boldsymbol{w}_k-\boldsymbol{w}_k^{(j)})\}). \label{surr4}
\end{align}
By substituting \eqref{surr0}-\eqref{surr4} into the objective function \eqref{sub_w}, the problem can be transformed into
\begin{subequations} \label{final_sub_w}
    \begin{align} \label{sub_w_f}
        \min_{\boldsymbol{w}_k} \quad &||\mathbf{F}\boldsymbol{w}_k||_2^2-\text{Re}\{\mathbf{f}^H\boldsymbol{w}_k\} \notag \\&+ \frac{1}{2\rho}\sum^3_{i=1}\Biggl(2\text{Re}\left\{\boldsymbol{w}_k^H(\tilde{\boldsymbol{\Theta}}_i-\mathbf{V})\boldsymbol{w}_k^{(j)}\right\} \notag \\
        &-4\hat{\lambda}_i\text{Re}\{||\boldsymbol{w}_k^{(j)}||^2(\boldsymbol{w}_k^{(j)})^H(\boldsymbol{w}_k-\boldsymbol{w}_k^{(j)})\} \notag \\
        &+|\boldsymbol{w}_k^H\text{vec}(\boldsymbol{\Theta}^H_i\mathbf{W}_k^{(j)})|^2+|\boldsymbol{w}_k^H\text{vec}(\boldsymbol{\Theta}_i\mathbf{W}_k^{(j)})|^2+o_i\Biggr) \\
\text{s.t.}\quad &-(\boldsymbol{w}_k^{(j)})^H\mathcal{G}_i^H\mathcal{G}_i\boldsymbol{w}_k^{(j)} \notag \\&+2\text{Re}\{(\boldsymbol{w}_k^{(j)})^H\mathcal{G}_i^H\mathcal{G}_i\boldsymbol{w}_k\} \geq \tilde{\Gamma}_t,~i \in \tilde{\mathcal{Q}}_k,\\
& ||\boldsymbol{w}_k||_2^2 \leq P_t,
    \end{align}
\end{subequations}
where $o_i, i=1,2,3$ collects all terms that do not depend on $\boldsymbol{w}_k$. Problem \eqref{sub_w_f} is a convex quadratically constrained quadratic programming (QCQP) problem, which can be solved optimally using existing optimization toolboxes such as CVX \cite{cvx}.

\begin{algorithm}[t]
    \caption{ADMM-Based BS Transmit Beamforming Algorithm for Solving Problem (P3.$k$)}
    \label{alg:ADMM}
    \KwIn{Optimized UA strategy: $\mathbf{U}^*$. Maximum iteration number: $N_{\text{iter}}$. Convergence threshold: $\varepsilon$.}
    \KwOut{BS transmit beamforming matrix $\mathbf{W}_k, ~\forall k\in\mathcal{K}$.}
    \textbf{Initialization:} Set feasible $\rho^{(0)} \geq 0$. Initialize $\boldsymbol{w}_k^{(0)}$ and $\boldsymbol{\Upsilon}^{(0)}$. Set $j=0$.\\
    \While{$j \leq N_{\text{iter}}$ and $\varepsilon^{(j)} > \varepsilon$}{
    \textbf{Outer loop:}\\
    \quad \textbf{Inner loop:}\\
        \qquad {\textbf{Step 1:} Update $\boldsymbol{\upsilon}^{(j)}$ and $\boldsymbol{b}^{(j)}$ based on \eqref{update_v} and \eqref{update_b}, respectively.}\\
        \qquad \textbf{Step 2:} Update $C^{(j)}$ and $\mathbf{q}^{(j)}$ via solving \eqref{sub_Cq}.\\
        \qquad \textbf{Step 3:} Update $\boldsymbol{w}_k^{(j)}$ given $\boldsymbol{\upsilon}^{(j)}, \boldsymbol{b}^{(j)}, C^{(j)}, \mathbf{q}^{(j)}$ via solving \eqref{final_sub_w}.\\
        \quad \textbf{End of Inner loop.}\\
        \quad \textbf{Step 4:} Update $\boldsymbol{\Upsilon}^{(j)}$ based on \eqref{update_R1}.\\
        \quad \textbf{Step 5:} Update $\rho^{(j+1)}$: $\rho^{(j+1)}=0.9\rho^{(j)}$.\\
        \quad \textbf{Step 6:} Update $j=j+1$.\\ 
        \quad \textbf{Step 7:} Calculate the increase of objective function as $\varepsilon^{(j)}$.\\
         \textbf{End of the Outer loop.}\\
    }
    Reshape $\boldsymbol{w}_k^*$ to $\mathbf{W}^*_k$. \\
    \Return{Optimized BS transmit beamforming matrix $\mathbf{W}^*_k, ~\forall k\in\mathcal{K}$.}
\end{algorithm}
d) Update Lagrangian dual
variables $\boldsymbol{\Upsilon}$:
The update of $\boldsymbol{\Upsilon}$ given $\boldsymbol{w}_k^{(j)},\boldsymbol{\upsilon}^{(j)},\boldsymbol{b}^{(j)},C^{(j)},\mathbf{q}^{(j)}$ obtained in the $j$-th iteration is given as \cite{AL}
\begin{equation} 
    \Upsilon_i = \Upsilon^{(j)}_i+\frac{1}{\rho^{(j)}}\left(Q_i(\mathbf{W}^{(j)}_k)-q^{(j)}_i\right), i=1,2,3, \label{update_R1}
\end{equation}
where $\Upsilon^{(j)}_i$ and $\rho^{(j)}$ denote the obtained dual variable and penalty factor in the $j$-th iteration, respectively. 
\subsubsection{Algorithm Summary}
The overall ADMM-based BS transmit beamforming algorithm for solving problem (P3.$k$) is summarized in Algorithm \ref{alg:ADMM}. The main complexity of Algorithm \ref{alg:ADMM} is due to solving \eqref{final_sub_w}. Given solution accuracy $\varepsilon$, the corresponding complexity for updating $\boldsymbol{w}_k$ via the interior-point method is in the order of $\mathcal{O}((M^{3.5}N^{3.5}+M^7)\rm{log}(1/\varepsilon))$ \cite{rangliu}. Noting that each step in the inner loop of Algorithm \ref{alg:ADMM} results in monotonically non-decreasing objective function values, the algorithm is guaranteed to converge to a stationary point because the search region of $\boldsymbol{w}_k$ is bounded and $\rho^{(j)}$ is shrunk in each iteration.

\begin{algorithm}[!ht]
    \caption{LLM-Enabled AO-Based Algorithm for Solving Problem (P1)}
    \label{alg:AO}
    \KwIn{Convergence threshold: $\xi$. Maximum iteration number: $N_{\rm{iter}}$.}
    \KwOut{UA strategy $\mathbf{U}$. BS transmit beamforming matrix $\mathbf{W}_k, ~\forall k$.}
    \textbf{Initialization:} Initialize $\mathbf{U}^{(0)}$ and $\mathbf{W}^{(0)}_k,~\forall k$. Set $j=0$.\\
    \While{$j \leq N_{\rm{iter}}$ and $ \xi^{(j)} > \xi$}
    {
    \textbf{Step 1:} Obtain $\mathbf{U}^{(j+1)}$ by running Algorithm \ref{algorithm1} with fixed $\mathbf{W}^{(j)}_k,~\forall k$.\\
    \textbf{Step 2:} \\
    \For{$k=1,2,\cdots, K$}{Obtain $\mathbf{W}^{(j+1)}_k$ by running Algorithm \ref{alg:ADMM} with fixed $\mathbf{U}^{(j+1)}$.}
    \textbf{Step 3:} Update $j=j+1$. \\
    \textbf{Step 4:} Calculate the increase of objective function value as $\xi^{(j)}$.\\
    }
     \Return{UA $\mathbf{U}^*$. The optimized BS transmit beamforming matrix $\mathbf{W}^*_k, ~\forall k\in\mathcal{K}$.} 
\end{algorithm}

\subsection{Overall Algorithm Summary}
The LLM-enabled AO-based algorithm design for solving problem (P1) is summarized in Algorithm \ref{alg:AO}. At the $j$-th iteration, we define the objective function of problem (P1) as $\mathcal{F}\left(
\{\mathbf{W}^{(j)}_k\}^K_{k=1},\mathbf{U}^{(j)}
\right)$. According to the AO-based algorithm, we have
\begin{align}
    \mathcal{F}\left(
\{\mathbf{W}^{(j)}_k\}^K_{k=1},\mathbf{U}^{(j)}
\right) &\stackrel{(a)}{\leq} \mathcal{F}\left(
\{\mathbf{W}^{(j)}_k\}^K_{k=1},\mathbf{U}^{(j+1)}
\right)\notag\\
&\stackrel{(b)}{\leq} \mathcal{F}\left(
\{\mathbf{W}^{(j+1)}_k\}^K_{k=1},\mathbf{U}^{(j+1)}
\right).
\end{align}
Note that inequality (a) holds since solution $\mathbf{U}^{(j+1)}$ to problem (P2) can be obtained for fixed $\{\mathbf{W}^{(j)}_k\}^K_{k=1}$. Inequality (b) holds since solution $\{\mathbf{W}^{(j+1)}_k\}^K_{k=1}$ to problem (P3) can be obtained for fixed $\mathbf{U}^{(j+1)}$. Thus, the update of variables leads to a non-decreasing objective function \eqref{eq:main_obj} after every iteration. Due to the power constraints and limited search space of the UA, problem (P1) has an upper bound. Thus, Algorithm \ref{alg:AO} is guaranteed to converge.

\section{Simulation Results}
\subsection{Simulation System Setup}
In this section, we present numerical results to verify the benefits of the proposed LLM-enabled AO-based algorithm.  In the considered multi-CU multi-BS ISAC network, we deploy $K$ ISAC BSs with $M$ transmit and receive antennas. For each ISAC BS $k \in \mathcal{K}$, there is one target $\tilde{Q}_k$ for detection and one target $\hat{Q}_k$ for parameter estimation in the associated target set $\mathcal{Q}_k$. The $K$ ISAC BSs serve $N$ ground CUs. We consider an area within 200 m $\times$ 200 m, where each BS is spaced between 80 and 160 m. Without loss of generality, we assume that all ground CUs are uniformly and randomly distributed, and the associated targets are randomly located at positions between 100 m and 160 m away from each ISAC BS. We assume that all the RCS of targets $\alpha_{t,i}, ~\forall i$ are randomly generated following CSCG distributions. We adopt the same initialization method for $\{\mathbf{W}_k\}^K_{k=1}$ as in \cite{rangliu} to maximize the sum power of the received signals of the targets and CUs using the Riemannian conjugate gradient (RCG) algorithm. The main system parameters are summarized in Table \ref{table1}. 
\begin{table}[h]
%\vspace{-1em}
    \centering
     \caption{Simulation Parameters}
    \label{table1}
    \renewcommand\arraystretch{1.5}
    \begin{tabular}{|c|c|}
        \hline
        Parameters & Values \\
        \hline
        The default number of ISAC BSs: $K$ & 3\\
        \hline
        The default number of ground CUs: $N$ & 10\\
        \hline
        The default number of antennas at each ISAC BS: $M$ & 24\\
        \hline
        The path loss factor at the reference distance $d_0=1$ m: $\beta_0$ & -30 dB\\
        \hline
        The path loss exponent of communication channels: $\varsigma_c$ & 2.4\\
        \hline
        The path loss exponent of radar channels: $\varsigma_t$ & 3.5\\
        \hline
        The Rician factor: $\kappa_{k,i},~\forall k,i$ & 3 dB\\
        \hline
        The communication and radar noise power: $\sigma^2_i,~\forall i,~\sigma^2_r$ & -90 dBm\\
        \hline
        The maximum transmit power at each BS: $P_t$ & 32 dBm\\
        \hline
        CRB of DoA estimation threshold: $\epsilon$ & 0.01 \\
        \hline
        Target detection threshold: $\Gamma_t$ & 7 dB \\
        \hline
        The number of collected samples: $L$ & 1024\\
        \hline
        The maximum number of iterations: $N_{\rm{iter}}$ & 100 \\
        \hline
    \end{tabular}
    %\vspace{-0.5em}
\end{table}
To evaluate the performance of our proposed LLM-enabled AO-based algorithm, we consider the following benchmark schemes for comparison:
\begin{figure*}[!h]
    \centering
    \subfigure[BS 1 ]{\includegraphics[scale=0.3]{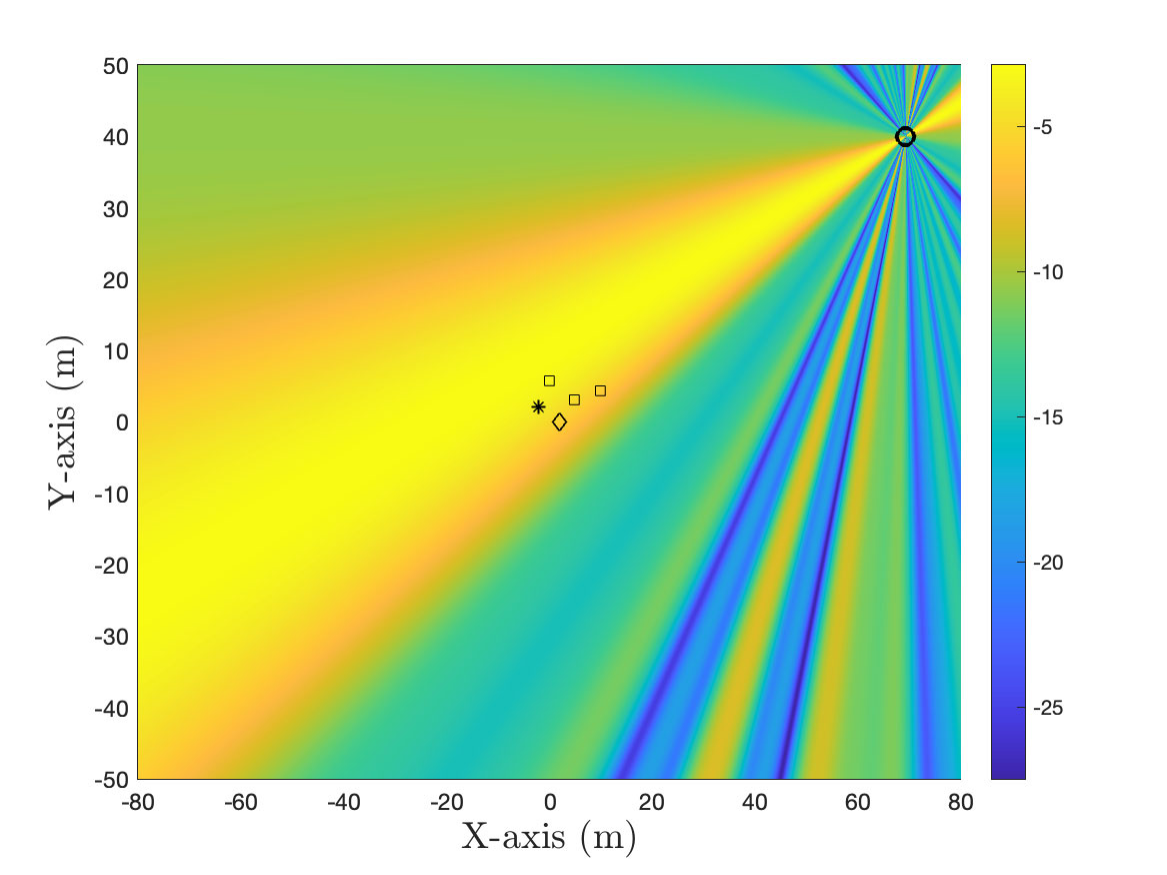}}
    \subfigure[BS 2]{\includegraphics[scale=0.3]{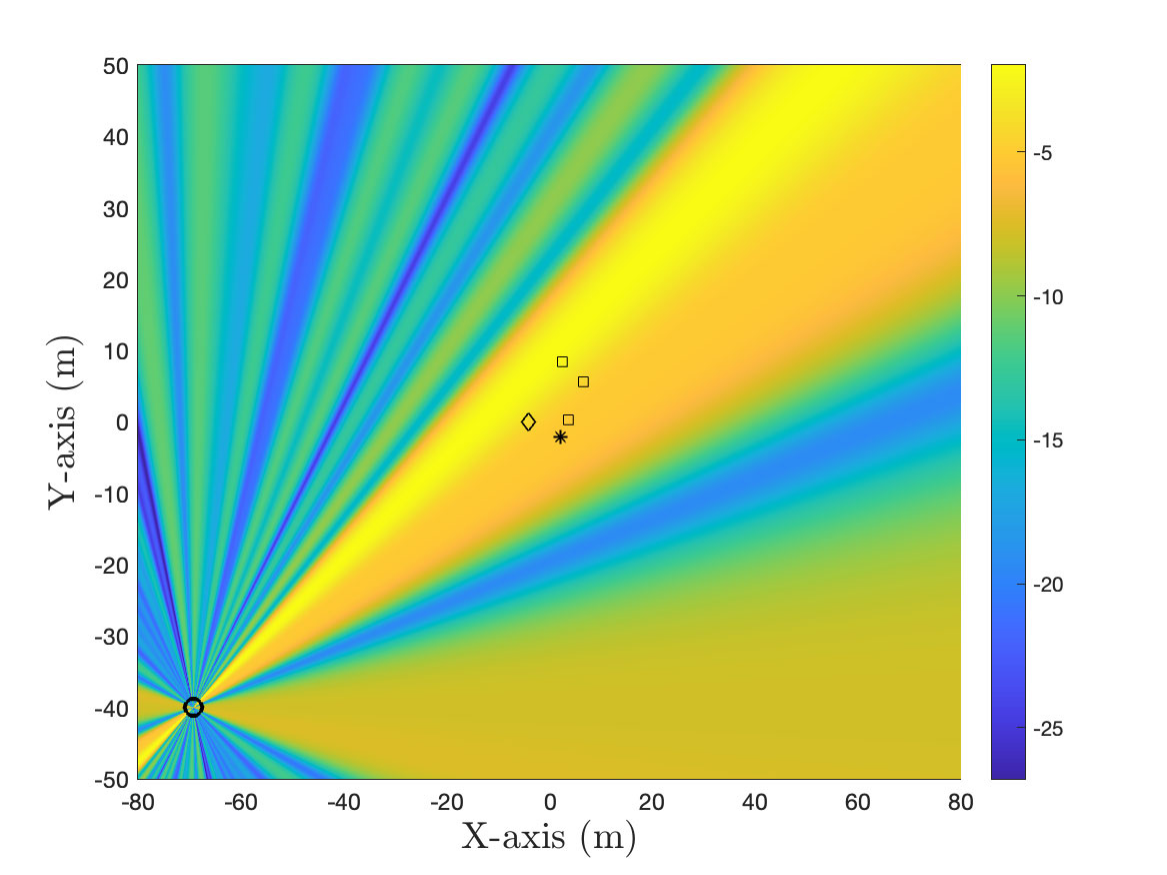}} 
    \subfigure[BS 3]{\includegraphics[scale=0.3]{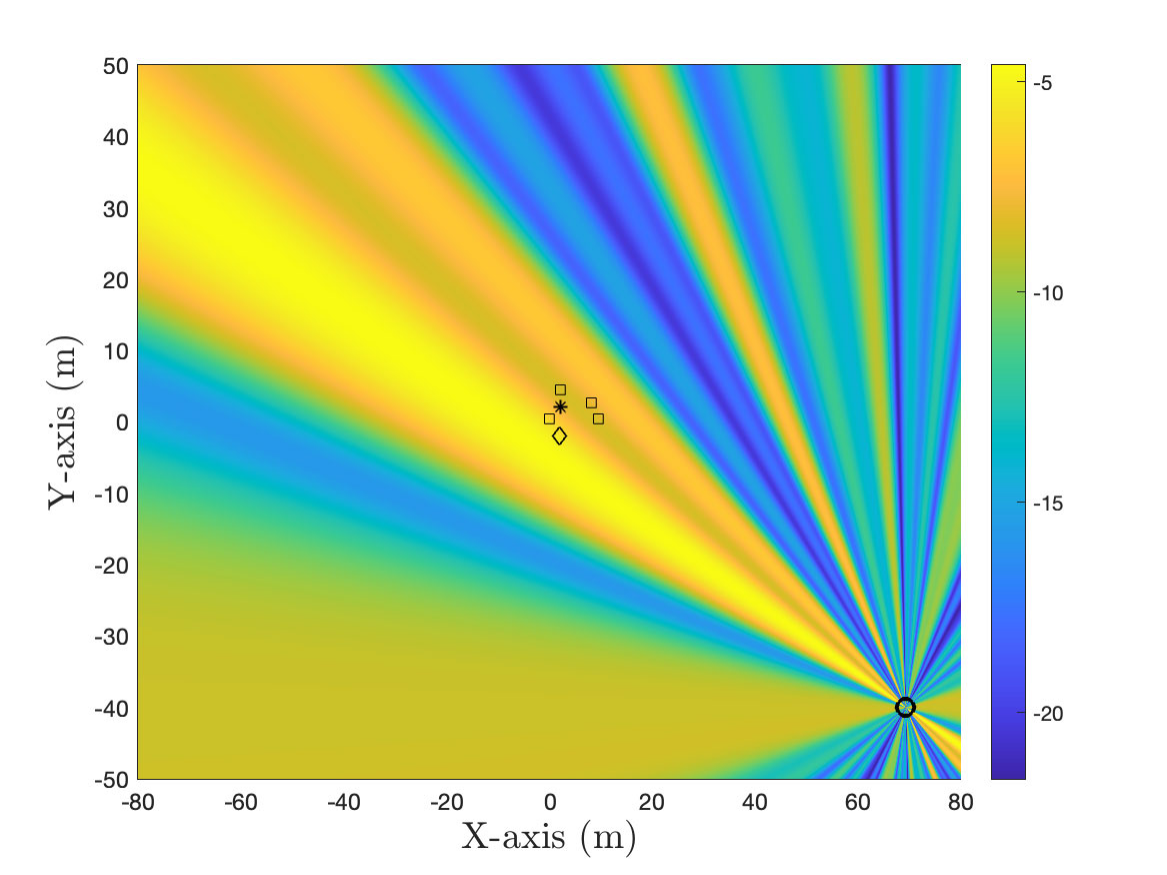}}
    \caption{Optimized transmit beampattern gains of different BSs with initialized UA (Beamforming only), where circle markers denote the BSs, the square markers denote the CUs, the star markers denote the target for parameter estimation, and the diamond markers denote the target for detection.}
    \label{fig:BP}
\end{figure*}
\begin{figure*}[!h]
    \centering
    \subfigure[BS 1 ]{\includegraphics[scale=0.3]{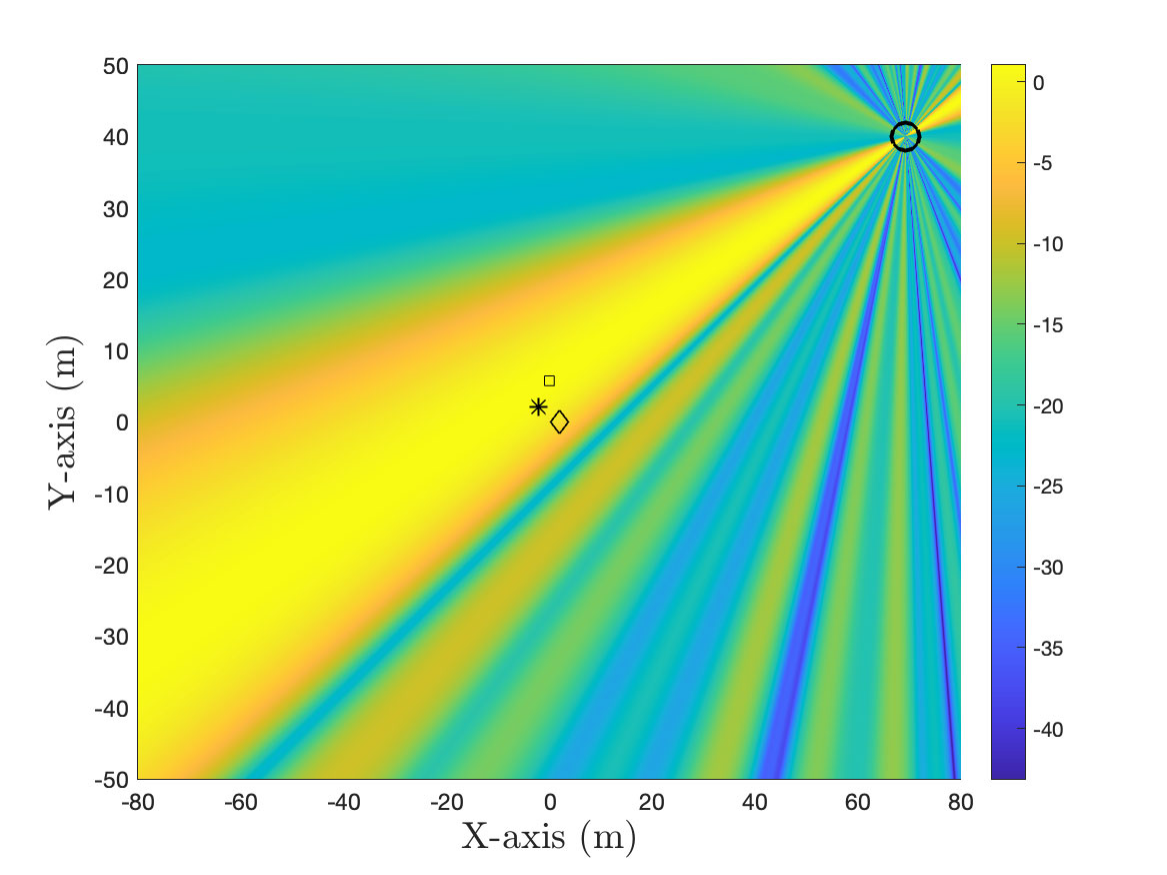}}
    \subfigure[BS 2]{\includegraphics[scale=0.3]{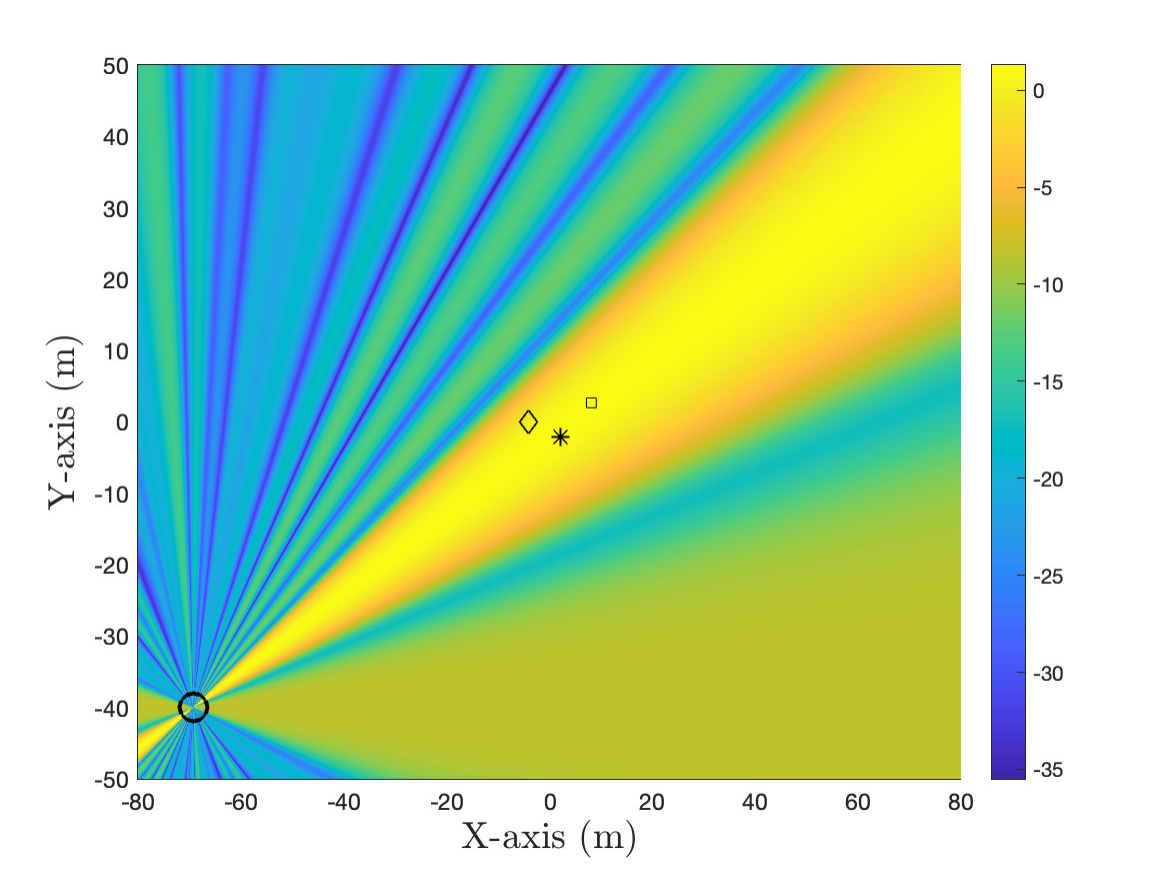}} 
    \subfigure[BS 3]{\includegraphics[scale=0.3]{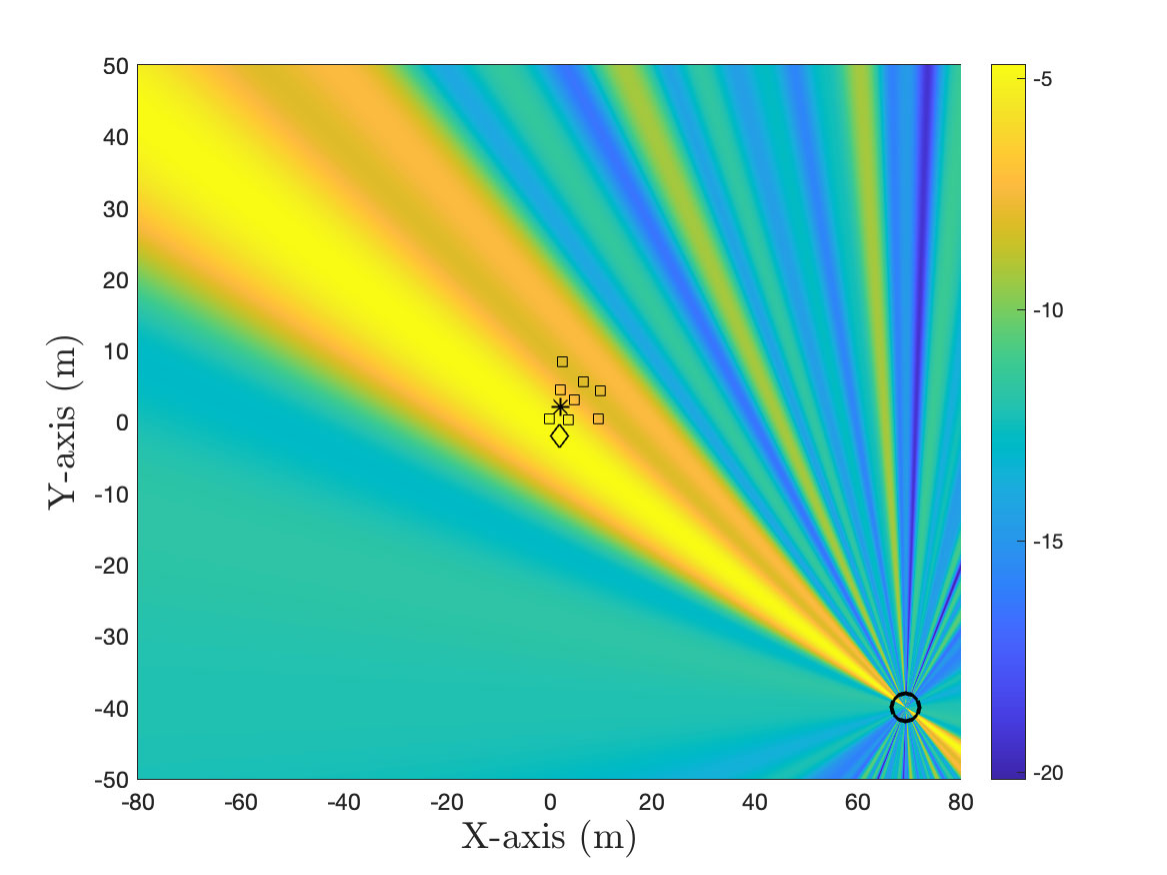}}
    \caption{Optimized transmit beampattern gains of different BSs with optimized UA (Proposed algorithm with the GPT-o1 model), where circle markers denote the BSs, the square markers denote the CUs, the star markers denote the target for parameter estimation, and the diamond markers denote the target for detection.}
    \label{fig:BP0}
\end{figure*}
\begin{itemize}
    \item Convex optimization with brute-force search method (Upper bound, \textbf{Convex plus BF}): This scheme utilizes the brute-force search to find the optimal UA in Step 1 of Algorithm \ref{alg:AO}, while optimizing the BS transmit beamforming with the Step 2 of Algorithm \ref{alg:AO}.
    \item Convex optimization with game theory (\textbf{Convex method only}): This scheme utilizes a low-complexity game theory-based method to optimize the UA in Step 1 of Algorithm \ref{alg:AO}. Specifically, this scheme first applies the Gale-Shapley algorithm \cite{gale-shapley} to find the initial matching between CUs and BSs. Then, considering that CUs may have an incentive for potential transfers from one heavily loaded base station (BS) to another lightly loaded BS, we formulate a coalition game based on \cite{coalition} to further optimize the user association (UA). The complexity of Step 1 is $\mathcal{O}(\mathcal{L}NK)$, where $\mathcal{L}$ is the number of iterations. Finally, the BS transmit beamforming is optimized using Step 2 of Algorithm \ref{alg:AO}.
    \item BS transmit beamforming optimization only (\textbf{Beamforming only}): This scheme only optimizes the BS transmit beamforming using Step 2 of Algorithm \ref{alg:AO}, while fixing the UA to the initial matching obtained by the Gale-Shapley algorithm.
    \item Proposed Algorithm using the GPT-o1 model (\textbf{Convex plus GPT-o1}): This scheme adopts the proposed LLM-enabled AO-based algorithm using the GPT-o1 model as the chosen LLM.
    \item Proposed Algorithm using the GPT-4-Turbo (\textbf{Convex plus GPT-4-Turbo}): This scheme adopts the proposed LLM-enabled AO-based algorithm using the GPT-4-Turbo model as the chosen LLM.
    \item Proposed Algorithm using the Claude 3.5 (\textbf{Convex plus Claude 3.5}): This scheme adopts the proposed LLM-enabled AO-based algorithm using the Claude 3.5 model as the chosen LLM.
    \item Proposed Algorithm using the Gemini 2.0 (\textbf{Convex plus Gemini 2.0}): This scheme adopts the proposed LLM-enabled AO-based algorithm using the Gemini 2.0 model as the chosen LLM.
\end{itemize}
In our simulations, we set the default parameter settings for all the LLMs used. All the simulation results were averaged over multiple independent runs. 
\subsection{Performance Evaluation}
\begin{figure}[h]
    \centering
    \includegraphics[scale=0.44]{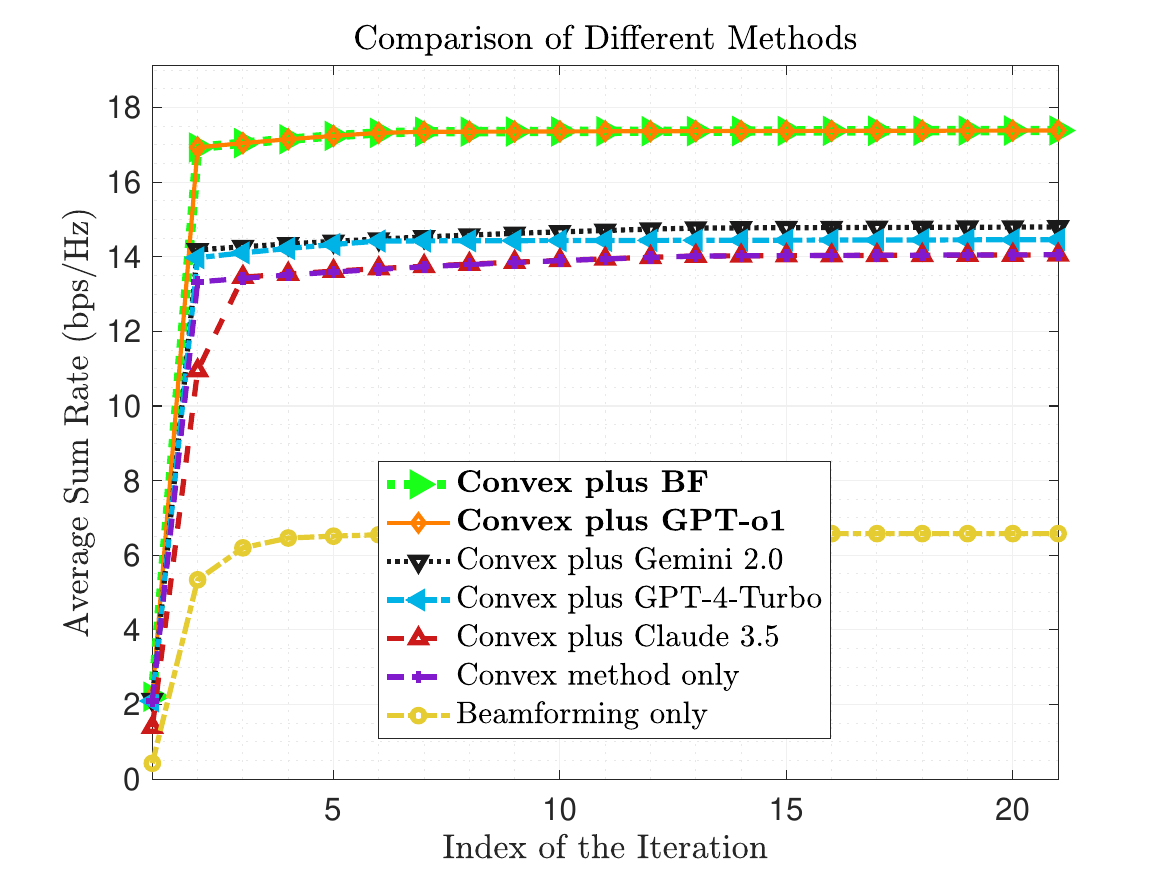}
    \caption{Convergence performance of the proposed algorithm.}
    \label{fig:Convergence}
\end{figure}
Fig. \ref{fig:BP} and Fig. \ref{fig:BP0} illustrate the optimized transmit beampattern gains of different BSs with the initial and optimized UA, respectively, where each BS senses the corresponding targets and serves the associated CUs. It can be observed that each BS generates the strongest beams towards the locations of the targets and associated CUs in both cases. Compared with the beamforming-only method shown in Fig. \ref{fig:BP}, the optimized beampattern gain of each BS from the proposed algorithm with the GPT-o1 model is much stronger and more concentrated towards the locations of the targets and CUs, while the beampattern gains towards other directions are reduced. Moreover, it is also observed that based on the proposed algorithm with the GPT-o1 model, BS 1 and BS 2 only serve a single CU, respectively, while BS 3 serves the rest of the CUs to maximize the overall communication performance. This demonstrates that our designed prompt engineering structure effectively guides the GPT-o1 model to find the optimal UA that maximizes the total sum rate, which also verifies the effectiveness of our proposed LLM-enabled AO-based algorithm.

In Fig. \ref{fig:Convergence}, we evaluate the convergence performance of the proposed algorithm compared with other benchmark schemes. The proposed algorithm with the GPT-o1 model converges within 10 iterations and achieves exactly the same performance as the Convex plus BF method, which achieves a performance upper bound. It is also observed that the proposed algorithm with the GPT-o1 model outperforms benchmark schemes using GPT-4-Turbo, Gemini 2.0, and Claude 3.5 in terms of the convergence speed. This can be attributed to the powerful reasoning capability of the GPT-o1 model. With the reasoning steps, GPT-o1 can clearly understand and efficiently solve the optimization problem by strictly following the chain of thought prompt. However, conversation models, including GPT-4-Turbo, Claude-3.5, and Gemini 2.0, are trained for general language conversations and have difficulty following the reasoning logic in the prompts and checking the correctness of their problem-solving logic. It can be concluded that the performance of the proposed algorithm is largely determined by the chosen LLM model, and the reasoning model GPT-o1 achieves the best performance in terms of convergence speed.

\begin{figure}[!htb]
    \centering
    \includegraphics[scale=0.44]{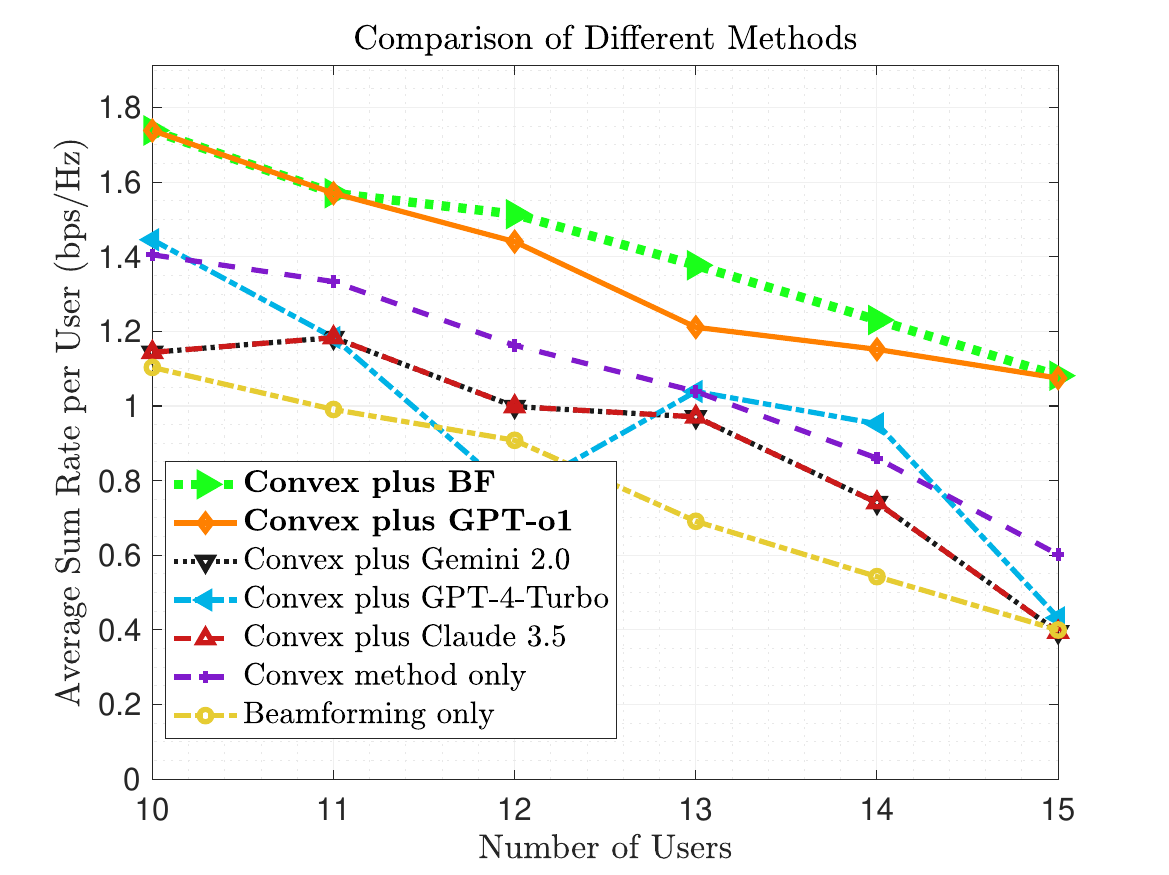}
    \caption{Average sum rate per CU as a function of the number of CUs for $M=24$ and $P_t=32~\rm{dBm}$.}
    \label{fig:users}
\end{figure}

In Fig. \ref{fig:users}, we evaluate the average sum rate per CU as a function of the number of CUs. As the number of CUs increases, the average sum rate per CU decreases since more CUs compete for the limited transmit power and bandwidth resources. It is observed that compared with other benchmark schemes, the proposed algorithm with the GPT-o1 model achieves the best performance in terms of the sum rate per CU, which is very close to the upper bound for a large number of CUs. To be more specific, it can be seen that the proposed algorithm with the GPT-o1 model can generate optimal results if a small number of CUs are considered. When the number of CUs increases, the GPT-o1 model can still generate near-optimal results with the exponential growth of the UA search space. However, compared with methods using LLMs, Convex method only scheme can generate smoother results with a clearer decreasing trend, which complies with the stability of convex-based optimization methods. Although the proposed algorithm with the GPT-4-Turbo model can sometimes outperform other benchmark methods, the generated results are not good enough and are sometimes even worse than the beamforming-only method. Similarly, Claude 3.5 and Gemini 2.0 achieve exactly the same performance, which is not as smooth as that of the Convex method only. First, this is because the sampling mechanisms of LLMs (controlled by, e.g., top-p and temperature) introduce randomness at each token, which yields diverse high-probability outputs in repeated runs rather than a single deterministic solution. Second, the fluctuating performance of Claude 3.5, Gemini 2.0, and GPT-4-Turbo is largely due to their general-purpose inference capabilities, which rely on broad statistical patterns rather than the precise step-by-step reasoning required for complex combinatorial optimization. In contrast, reasoning models like GPT-o1 are fine-tuned with dedicated reasoning capabilities that better navigate the exponential search space, resulting in more deterministic outputs. Hence, it can be concluded that the performance of the proposed algorithm is largely determined by the chosen LLM model, and the reasoning GPT-o1 model is more likely to yield deterministic and high-quality solutions compared with conversation models.
\begin{figure}[!htb]
    \centering
    \includegraphics[scale=0.44]{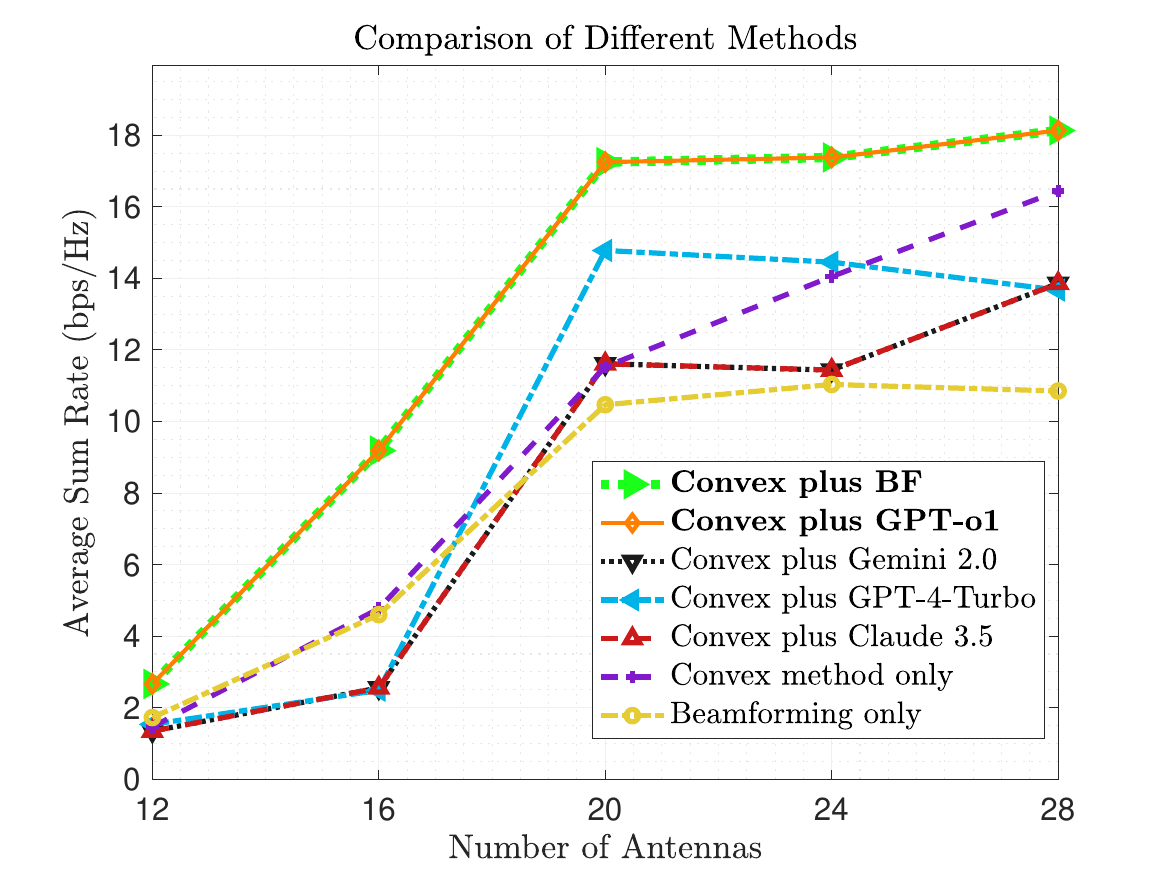}
    \caption{Average sum rate as a function of the number of antennas for $N=10$ and $P_t=32 ~\rm{dBm}$.}
    \label{fig:antennas}
\end{figure}

In Fig. \ref{fig:antennas}, we evaluate the average sum rate as a function of the number of antennas. We observe a small performance gap between the proposed algorithm and other benchmark methods when the number of antennas is small. This is because there are not enough DoFs for beamforming to sense targets and serve CUs.  As can be observed, as the number of antennas increases, the average sum rate increases since additional antennas can provide more spatial DoFs for more efficient beamforming. It is also observed that the proposed algorithm with the GPT-o1 model achieves the best performance in terms of the sum rate and approaches the upper bound. Besides, the proposed algorithm with conversation LLMs still fails to generate the expected results due to the random output of such LLMs.

\begin{figure}[!htb]
    \centering
    \includegraphics[scale=0.44]{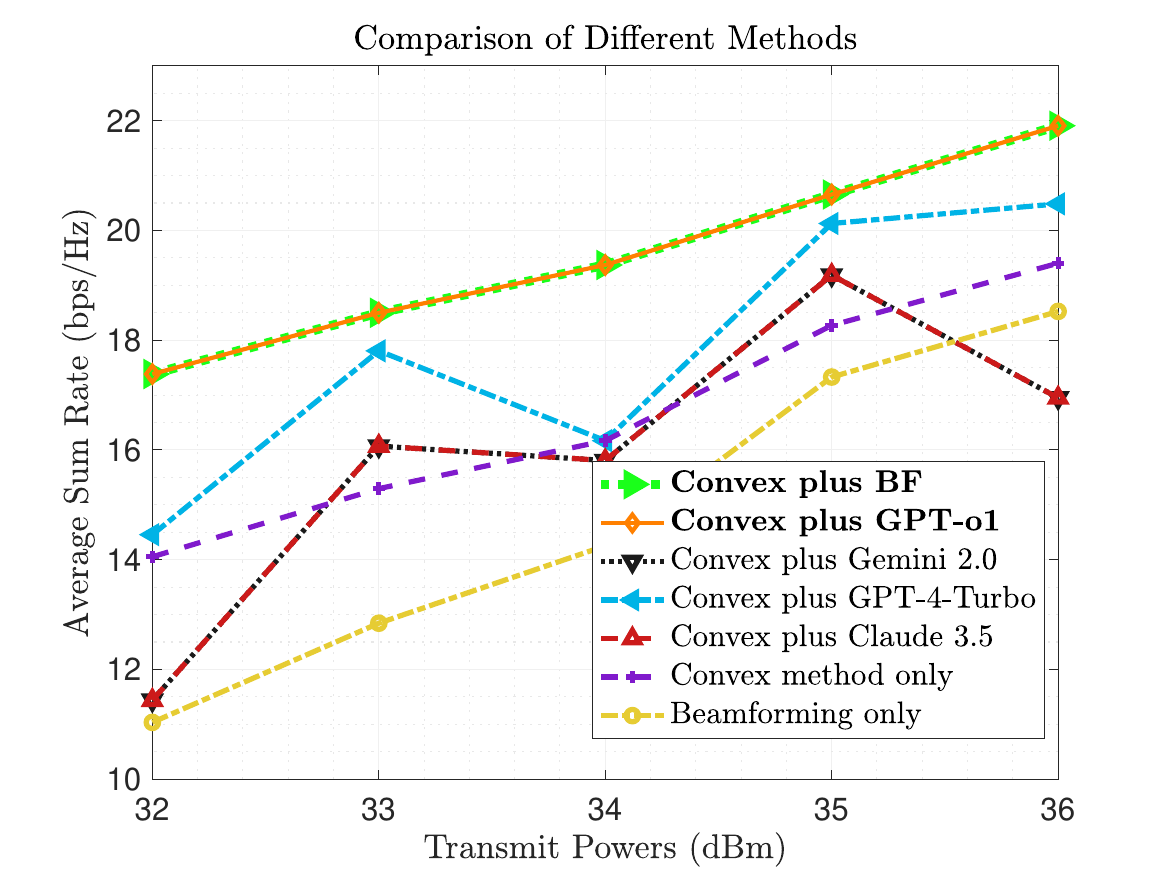}
    \caption{Averaged sum rate as a function of the transmit power for $N=10$ and $M=24$.}
    \label{fig:powers}
\end{figure}

In Fig. \ref{fig:powers}, we evaluate the average sum rate as a function of the transmit power. It is observed that as the transmit power increases, the average sum rate also increases because more power can be allocated to each CU for better communication performance, which is an obvious and predictable result. In addition, the proposed algorithm achieves the upper bound while outperforming other benchmark methods. Compared with the GPT-o1 model, although the use of conversation LLMs can sometimes outperform the Convex method only scheme, such LLMs still fail to generate deterministic and smooth results due to their stochastic generation process and limited capabilities in handling complex optimization problems.

\begin{figure}[!htb]
    \centering
    \includegraphics[scale=0.44]{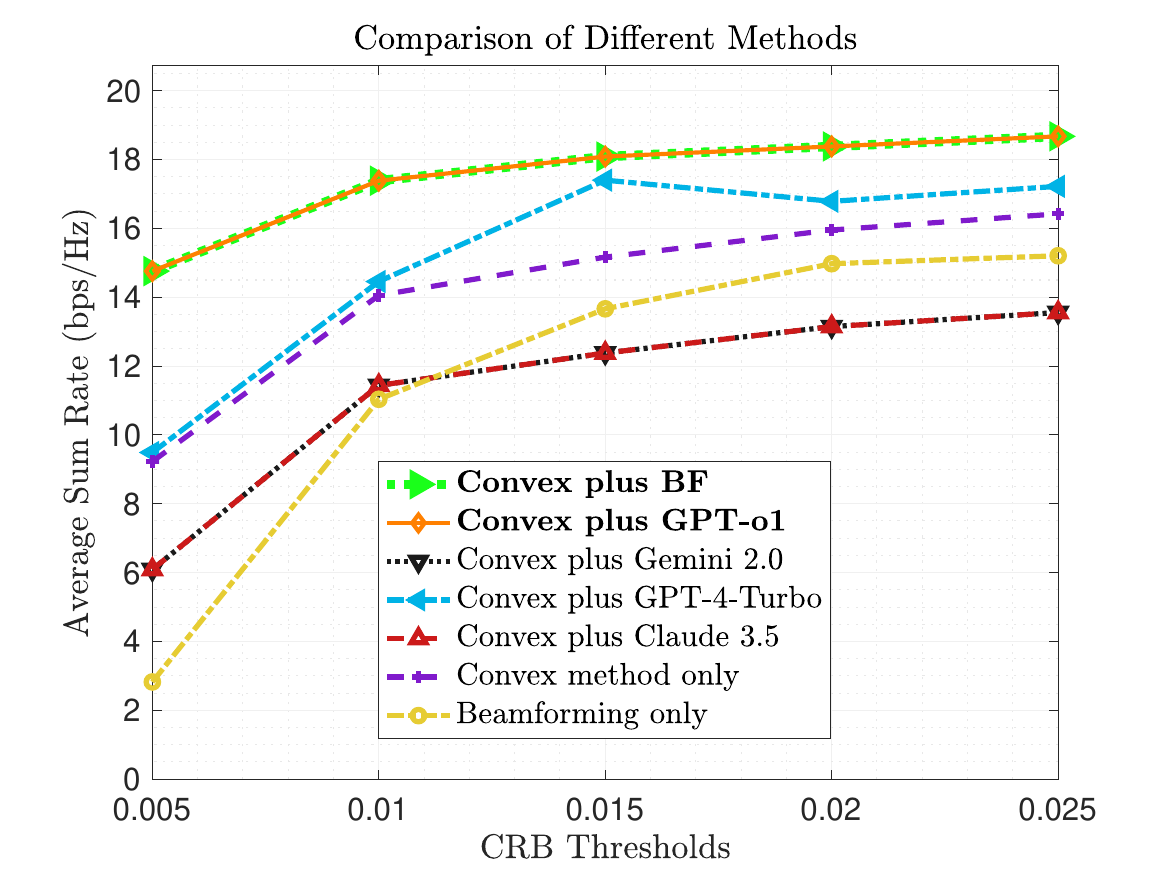}
    \caption{Average sum rate as a function of the values of CRB thresholds for $N=10$, $M=24$, and $P_t=32 ~\rm{dBm}$}
    \label{fig:crbs}
\end{figure}

In Fig. \ref{fig:crbs}, we evaluate the average sum rate as a function of the values of CRB thresholds. We can observe a high-performance gap for all methods from the CRB threshold of 0.005 to that of 0.01. This is because such a CRB threshold of 0.005 is so tight that the BSs are forced to allocate most of the transmit power for sensing, thus compromising the communication performance. There is also a clear trade-off between communication and sensing. As the CRB thresholds increase, the average sum rate also increases since the sensing requirements become looser, and more power can be allocated for communications. In addition, the proposed algorithm with the GPT-o1 model reaches the upper bound while outperforming other benchmark methods. The proposed algorithm with Gemini 2.0 and Claude 3.5 even generates results worse than the Beamforming only scheme, which again verifies the benefits of using reasoning LLMs for the performance of our proposed algorithm.

\section{Conclusions}
We considered a multi-BS multi-CU ISAC system and proposed an LLM-enabled AO-based algorithm to jointly optimize the multi-BS transmit beamforming and UA strategy to maximize the total communication sum rate while ensuring the sensing requirements. Firstly, we formulated the corresponding optimization problem based on appropriate performance metrics, including the radar SNR for sensing, CRB for parameter estimation, and downlink sum rate for communications. Secondly, we decomposed the original problem into two sub-problems, namely the UA optimization and beamforming optimization problems. Thirdly, we proposed integrating LLMs and convex-based optimization into an LLM-enabled AO-based algorithm framework to solve the two sub-problems iteratively. LLMs learn to efficiently optimize the UA given expert knowledge through in-context, few-shot, chain of thought, and self-reflection prompt engineering, which demonstrates the strong and flexible adaptability of LLMs without extensive training. Then, the convex-based optimization is used to handle beamforming optimization based on the FP, MM, and ADMM techniques. Finally, our numerical results demonstrated that the algorithm proposed with the GPT-o1 model achieves a performance close to the upper bound and outperforms the Convex method only scheme and other LLM models (GPT-4-Turbo, Claude 3.5, Gemini 2.0) in terms of total sum rate and convergence speed, verifying the effectiveness of integrating reasoning LLMs with the convex-based optimization algorithm framework by combining both of their benefits.

 \begin{appendices}  \section{}
      \label{appenA}
      Based on the derivation of FIM in \eqref{FIM1}, we first calculate the partial derivatives of $\boldsymbol{\mu}(\boldsymbol{\xi}_{k,i})$ with respect to each parameter in $\boldsymbol{\xi}_{k,i}$, which yields
      \begin{equation}
          \frac{\partial\boldsymbol{\mu}(\boldsymbol{\xi}_{k,i})}{\partial\phi_{k,i}}=\left[\begin{array}{c}
\alpha_{t,i}\dot{\mathbf{G}}(\phi_{k,i})\mathbf{W}_k\mathbf{s}_k[1] \\
\vdots \\
\alpha_{t,i}\dot{\mathbf{G}}(\phi_{k,i})\mathbf{W}_k\mathbf{s}_k[L]
\end{array}\right] \in \mathbb{C}^{ML\times1},
\end{equation}
\begin{equation}
         \frac{\partial\boldsymbol{\mu}(\boldsymbol{\xi}_{k,i})}{\partial\boldsymbol{\alpha}_i^T}=[1~\mathfrak{j}] \otimes \left[\begin{array}{c}
             {\mathbf{G}}(\phi_{k,i})\mathbf{W}_k\mathbf{s}_k[1]  \\
             \vdots\\
             {\mathbf{G}}(\phi_{k,i})\mathbf{W}_k\mathbf{s}_k[L]
         \end{array}\right], \in \mathbb{C}^{ML\times2},
\end{equation}
where $\dot{\mathbf{G}}(\phi_{k,i})$ denotes the partial derivative of $\mathbf{G}(\phi_{k,i})$ with respect to $\theta_{k,i}$. The elements of the FIM in \eqref{FIM2} are given by
\begin{align}
    &{J}(\boldsymbol{\xi}_{k,i})_{\phi_{k,i}\phi_{k,i}} \notag \\
    &=\frac{2}{\sigma^2_r}\sum_{l=1}^{L}\bar{\alpha}_{t,i}\mathbf{s}^H_k[l]\mathbf{W}^H_k\dot{\mathbf{G}}^H(\phi_{k,i})\alpha_{t,i}\dot{\mathbf{G}}(\phi_{k,i})\mathbf{W}_k\mathbf{s}_k[l] \notag \\
    &=\frac{2}{\sigma^2_r}|\alpha_{t,i}|^2\text{Tr}\{\dot{\mathbf{G}}(\phi_{k,i})\mathbf{W}_k\sum_{l=1}^L\mathbf{s}_k[l]\mathbf{s}^H_k[l]\mathbf{W}^h_k\dot{\mathbf{G}}^H(\phi_{k,i})\} \notag \\
    &=\frac{2L|\alpha_{t,i}|^2}{\sigma^2_r}\text{Tr}\{\dot{\mathbf{G}}^H(\phi_{k,i})\dot{\mathbf{G}}(\phi_{k,i})\mathbf{W}_k\mathbf{W}^H_k\},
\end{align}
where we assume $\frac{1}{L}\sum_{l=1}^L\mathbf{s}_k[l]\mathbf{s}^H_k[l]\approx\mathbf{I}_{N+M}$ for sufficiently large $L$.

Similarly, we have
\begin{align}
    &\mathbf{J}(\boldsymbol{\xi}_{k,i})_{\phi_{k,i}\boldsymbol{\alpha}_i^T} \notag \\
    &=\text{Re}\bigg\{\frac{2}{\sigma^2_r}\bigg[\sum_{l=1}^{L}\bar{\alpha}_{t,i}\mathbf{s}^H_k[l]\mathbf{W}^H_k\dot{\mathbf{G}}^H(\phi_{k,i}){\mathbf{G}}(\phi_{k,i})\mathbf{W}_k\mathbf{s}_k[l], \notag \\
    &\quad \sum_{l=1}^{L}\bar{\alpha}_{t,i}\mathbf{s}^H_k[l]\mathbf{W}^H_k\dot{\mathbf{G}}^H(\phi_{k,i}){\mathbf{G}}(\phi_{k,i})\mathbf{W}_k\mathbf{s}_k[l]\times \mathfrak{j}\bigg]\bigg\}, \notag \\
    &=\text{Re}\bigg\{\frac{2L\bar{\alpha}_{k,i}}{\sigma^2_r}\text{Tr}\{\dot{\mathbf{G}}^H(\phi_{k,i}){\mathbf{G}}(\phi_{k,i})\mathbf{W}_k\mathbf{W}^H_k\} [1~\mathfrak{j}]\bigg\}.
\end{align}
\begin{align}
    &\mathbf{J}(\boldsymbol{\xi}_{k,i})_{\boldsymbol{\alpha}_i\boldsymbol{\alpha}_i^T} \notag \\
    &=\text{Re}\bigg\{\frac{2L}{\sigma^2_r}\sum^L_{l=1}\left[\begin{array}{cc}
      1   & \mathfrak{j} \\
      \mathfrak{-j}   & 1
    \end{array}\right] \notag \\&\times\mathbf{s}^H_k[l]\mathbf{W}^H_k{\mathbf{G}}^H(\phi_{k,i}){\mathbf{G}}(\phi_{k,i})\mathbf{W}_k\mathbf{s}_k[l]\bigg\}, \notag \\
    &=\frac{2L}{\sigma^2_r}\text{Re}\bigg\{\text{Tr}\{{\mathbf{G}}^H(\phi_{k,i}){\mathbf{G}}(\phi_{k,i})\mathbf{W}_k\mathbf{W}^H_k\}\left[\begin{array}{cc}
      1   & \mathfrak{j} \\
      \mathfrak{-j}   & 1
    \end{array}\right]\bigg\}, \notag \\
    &=\frac{2L}{\sigma^2_r}\text{Tr}\{{\mathbf{G}}^H(\phi_{k,i}){\mathbf{G}}(\phi_{k,i})\mathbf{W}_k\mathbf{W}^H_k\}\mathbf{I}_2.
\end{align}

To rearrange the FIM in \eqref{FIM2}, based on the deriviations above, we define $Q_j(\cdot)$ as
\begin{equation}
    Q_j(\mathbf{W}_k) \triangleq \text{Tr}\{\boldsymbol{\Theta}_j\mathbf{W}_k\mathbf{W}^H_k\}, ~j=1,2,3,
\end{equation}
where 
\begin{align}
    \boldsymbol{\Theta}_1&\triangleq\frac{2L|\alpha_{t,i}|^2}{\sigma^2_r}\dot{\mathbf{G}}^H(\phi_{k,i})\dot{\mathbf{G}}(\phi_{k,i}),\\
    \boldsymbol{\Theta}_2&\triangleq\frac{2L\bar{\alpha}_{t,i}}{\sigma^2_r}\dot{\mathbf{G}}^H(\phi_{k,i}){\mathbf{G}}(\phi_{k,i}),\\
    \boldsymbol{\Theta}_3&\triangleq\frac{2L}{\sigma^2_r}{\mathbf{G}}^H(\phi_{k,i}){\mathbf{G}}(\phi_{k,i}).
\end{align}
  \end{appendices}


\begin{thebibliography}{1}
\bibitem{isac_survey0}
F. Liu, C. Masouros, A. P. Petropulu, H. Griffiths, and L. Hanzo, ``Joint radar and communication design: Applications, state-of-the-art, and the road ahead,'' \textit{IEEE Trans. Commun.}, vol. 68, no. 6, pp. 3834--3862, Jun. 2020.
\bibitem{survey0}
Y. Xiao, Z. Ye, M. Wu, H. Li, M. Xiao, M. S. Alouini, A. Al-Hourani, and S. Cioni, ``Space-air-ground integrated wireless networks for 6G: Basics, key technologies, and future trends," \textit{IEEE J. Sel. Areas Commun.}, vol. 42, no. 12, pp. 3327--3354, Dec. 2024.
\bibitem{isac_survey1}
F. Liu, Y. Cui, C. Masouros, J. Xu, T. Han, Y. Eldar, and S. Buzzi, ``Integrated Sensing and Communication: Towards dual-
functional wireless networks for 6G and beyond,'' \textit{IEEE J. Sel. Areas
Commun.}, vol. 40, no. 6, pp. 1728--1767, Jun. 2022.
\bibitem{isac0}
C. Sturm and W. Wiesbeck, ``Waveform design and signal processing
aspects for fusion of wireless communications and radar sensing,'' \textit{Proc.
IEEE}, vol. 99, no. 7, pp. 1236--1259, Jul. 2011.
\bibitem{isac01}
P. Kumari, A. Mezghani, and R. W. Heath, ``JCR70: A low-complexity
millimeter-wave proof-of-concept platform for a fully-digital SIMO
joint communication-radar,'' \textit{IEEE Open J. Veh. Technol.}, vol. 2,
pp. 218--234, 2021.
\bibitem{MIMO0}
A. J. Paulraj and T. Kailath, ``Increasing capacity in wireless broadcast
systems using distributed transmission/directional reception (DTDR),''
U.S. Patent 5 345 599, Sep. 6, 1994.
\bibitem{MIMO1}
D. Cohen, D. Cohen, and Y. C. Eldar, ``High resolution FDMA
MIMO radar,'' \textit{IEEE Trans. Aerosp. Electron. Syst.}, vol. 56, no. 4,
pp. 2806--2822, Aug. 2020.
\bibitem{MIMO00}
S. Ye, M. Xiao, M.W. Kwan, Z. Ma, Y. Huang, G. Karagiannidis, and P. Fan, ``Extremely large aperture array (ELAA) communications: Foundations, research advances and challenges," \textit{IEEE Open Journal of the Commun. Soc.}, vol. 5, pp. 7075--7120, Oct. 2024.
\bibitem{Wang}
Z. Wang, X. Mu, and Y. Liu, ``STARS enabled integrated sensing and communication,'' \textit{IEEE Trans. Wireless Commun.}, vol. 22, no. 10, pp. 6750--6765, Oct. 2023.
\bibitem{MIMO2}
Z. Lyu, G. Zhu, and J. Xu, ``Joint maneuver and beamforming design
for UAV-enabled integrated sensing and communication,'' \textit{IEEE Trans.
Wireless Commun.}, vol. 22, no. 4, pp. 2424--2440, Apr. 2023.
\bibitem{MIMO2_1}
D. Deng, W. Zhou, X. Li, D. B. da Costa, D. W. K. Ng, and A. Nallanathan, ``Joint beamforming and UAV trajectory optimization for covert communications in ISAC networks'', {\it{IEEE Trans.
Wireless Commun.}}, vol.24, no.2, pp. 1016--1030, Feb. 2025.
\bibitem{MIMO3}
F. Liu, C. Masouros, A. Li, H. Sun, and L. Hanzo, ``MU-MIMO
communications with MIMO radar: From co-existence to joint transmission,'' \textit{IEEE Trans. Wireless Commun.}, vol. 17, no. 4, pp. 2755--2770, Apr. 2018.
\bibitem{MIMO4}
X. Liu, T. Huang, N. Shlezinger, Y. Liu, J. Zhou, and Y. C. Eldar, ``Joint
transmit beamforming for multiuser MIMO communications and MIMO
radar,'' \textit{IEEE Trans. Signal Process.}, vol. 68, pp. 3929--3944, 2020.
\bibitem{MIMO5}
W. Mao, Y. Lu, C. Y. Chi, B. Ai, Z. Zhong, and Z. Ding, ``Communication-sensing region for cell-free massive MIMO ISAC systems,'' \textit{IEEE Trans. Wireless Commun.}, vol. 23, no. 9, pp. 12396--12411, Sept. 2024.
\bibitem{rangliu}
R. Liu, M. Li, Q. Liu, and A. Lee Swindlehurst, ``SNR/CRB-constrained
joint beamforming and reflection designs for RIS-ISAC systems,'' \textit{IEEE
Trans. Wireless Commun.}, vol. 23, no. 7, pp. 7456--7470, Jul. 2024.
\bibitem{MIMO6}
Q. Zhu, M. Li, R. Liu, and Q. Liu, ``Cramér-Rao bound optimization for active RIS-empowered ISAC systems,'' \textit{IEEE Trans. Wireless Commun.}, vol. 23, no. 9, pp. 11723--11736, Sept. 2024.
\bibitem{MIMO7}
C. Liu, W. Yuan, S. Li, X. Liu, H. Li, and D. W. K. Ng, ``Learning-based predictive beamforming for integrated sensing and communication in vehicular networks," \textit{IEEE J. Sel. Areas Commun.}, vol. 40, no. 8, pp. 2317--2334, Aug. 2022.
\bibitem{MIMO8}
X. Zhang, W. Yuan, C. Liu, J. Wu, and D. W. K. Ng, ``Predictive beamforming for vehicles with complex behaviors in ISAC systems: A deep learning approach," \textit{IEEE J. Sel. Top. Signal Process.}, vol. 18, no. 5, pp. 828--841, Jul. 2024.
\bibitem{Jiang}
W. Jiang, D. Ma, Z. Wei, Z. Feng, P. Zhang, and J. Peng, ``ISAC-NET: Model-driven deep learning for integrated passive sensing and communication,'' \textit{IEEE Trans. Wireless Commun.} vol. 72, no. 8, pp. 4692--4707, Aug. 2024.
\bibitem{RL1}
C. Wang, G. Li, H. Zhang, K. K. Wong, Z. Li, and D, W. K. Ng, ``Fluid antenna system liberating multiuser MIMO for ISAC via deep reinforcement learning," \textit{IEEE Trans. Wireless Commun.}, vol. 23, no. 9, pp. 10879--10894, Sept. 2024.
\bibitem{RL2}
P. Saikia, K. Singh, W. J. Huang, and T. Q. Duong, ``Hybrid deep reinforcement learning for enhancing localization and communication efficiency in RIS-aided cooperative ISAC systems," \textit{IEEE Internet Things J}, vol. 11, no. 18, pp. 29494--29510, Sept. 2024.
\bibitem{RL3}
L. Yang, Y. Wei, Z. Feng, Q. Zhang, and Z. Han, ``Deep reinforcement learning-based resource allocation for integrated sensing, communication, and computation in vehicular network,'' \textit{IEEE Trans. Wireless Commun.}, vol. 23, no. 12, pp. 18608--18622, Dec. 2024.
\bibitem{SDR}
Z. Q. Luo, W. K. Ma, A. M. So, Y. Ye, and S. Zhang, ``Semidefinite relaxation of quadratic optimization problems,'' \textit{IEEE Signal Process. Mag.}, vol. 27, no. 3, pp. 20--34, May 2010.
\bibitem{DL}
Y. LeCun, Y. Bengio, and G. Hinton, ``Deep learning,'' \textit{Nature}, vol. 521, no. 7553, pp. 436--444, Nov. 2015.
\bibitem{MOPsurvey}
Z. Fei, B. Li, S. Yang, C. Xing, H. Chen, and L. Hanzo, ``A survey of multi-objective optimization in wireless sensor networks: Metrics, algorithms, and open problems,'' {\it{IEEE Commun. Surveys Tuts.}}, vol. 19, no. 1, pp. 550--586, 1st Quart., 2017.
\bibitem{Sun}
R. Sun, N. Cheng, C. Li, F. Chen, and W. Chen, ``Knowledge-driven deep learning paradigms for wireless network optimization in 6G,'' \textit{IEEE Netw.}, vol. 38, no. 2, pp. 70--78, Mar. 2024.
\bibitem{LLM01}
Y. Xu et al., ``ChatGLM-Math: Improving math problem-solving in large language models with a self-critique pipeline'',  in \textit{Proc. Conf. Empir. Methods Nat. Lang. Process. (EMNLP)}, pp. 9733--9760, Nov. 2024.
\bibitem{LLM02}
C. Yang, X. Wang, Y. Lu, H. Liu, Q. V. Le, D. Zhou, and X. Chen, ``Large language models as optimizers'', in \textit{Proc. Int. Conf. Learn. Representations (ICLR)}, pp. 1--22, Jan. 2024. 
\bibitem{LLM03}
F. Liu, X. Lin, S. Yao, Z. Wang, X. Tong, M. Yuan, and Q. Zhang,
``Large language model for multiobjective evolutionary optimization'', in \textit{Proc. Int. Conf. Evolutionary Multi-Criterion Optimization (EMO)}, pp. 178--191, Feb. 2025.
\bibitem{LLM05}
T. Liu, N. Astorga, N. Seedat, and M. Schaar, ``Large language models to enhance bayesian optimization'', in \textit{Proc. Int. Conf. Learn. Representations (ICLR)}, pp. 1--33, Jan. 2024.
\bibitem{LLM06}
T. Bömer, N. Koltermann, M. Disselnmeyer, L. Dörr, and A. Meyer, ``Leveraging large language models to develop heuristics for emerging optimization problems'', 2025. [Online]. Available: https://arxiv.org/abs/2503.03350.
\bibitem{LLM07}
M. Pluhacek, A. Kazikova, T. Kadavy, A. Viktorin, and R. Senkerik,
“Leveraging large language models for the generation of novel metaheuristic optimization algorithms,” in \textit{Proc. Companion Conf. Genetic and Evolutionary Computation Conf. (GECCO)}, pp. 1812--1820, Jul. 2023.
\bibitem{LLMsurvey}
W. X. Zhao et al., ``A survey of large language
models,'' 2023. [Online]. Available: https://arxiv.org/abs/2303.18223.
\bibitem{LLM22}
H. Zhou et al., ``Large language model (LLM) for telecommunications: A comprehensive survey on principles, key techniques, and opportunities," \textit{IEEE Commun. Surv. Tutorials.}, Early Access, 2024.
% \bibitem{LLM33}
% S. Javaid, H. Fahim, B. He and N. Saeed, ``Large language models for UAVs: Current state and pathways to the future," \textit{IEEE Open Journal of the Veh. Techn. Society}, vol. 5, pp. 1166--1192, 2024.
\bibitem{WirelessLLM}
J. Shao et al., ``WirelessLLM: Empowering large language models towards wireless intelligence," \textit{J. Commun. Inf. Netw.}, vol. 9, no. 2, pp. 99--112, Jun. 2024.
\bibitem{LLM4}
J. Tong, J. Shao, Q. Wu, W. Guo, Z. Li, Z. Lin, and J. Zhang, ``WirelessAgent: Large language model agents for intelligent wireless networks,” 2024, [Online]. Available: https://arxiv.org/abs/2409.07964.
\bibitem{LLM5}
H. Quan, W. Ni, T. Zhang, X. Ye, Z. Xie, S. Wang, Y. Liu, and H. Song, ``Large language model agents for radio map generation and wireless network planning,'' \textit{IEEE Networking Letters}, Early Access, 2025.
\bibitem{LLM6}
K. Qiu, S. Bakirtzis, I. Wassell, H. Song, J. Zhang, and K. Wang, ``Large language model-based wireless network design,''  \textit{IEEE Wireless Commun. Lett.}, vol. 13, no. 12, pp. 3340--3344, Dec. 2024.
\bibitem{LLM7}
H. Xie, Z. Qin, X. Tao, and Z. Han, ``Toward intelligent communications: Large model empowered semantic communications," \textit{IEEE Commun. Mag.}, vol. 63, no. 1, pp. 69--75, Jan. 2025.
\bibitem{LLM8}
Y. Shen, J. Shao, X. Zhang, Z. Lin, H. Pan, D. Li, J. Zhang, and KB. Letaief, ``Large language models empowered autonomous edge AI for connected intelligence,'' \textit{IEEE Commun. Mag.}, vol. 62, no. 10, pp. 140--146, Oct. 2024.
\bibitem{LLM9}
H. Li, M. Xiao, K. Wang, D. I. Kim, and M. Debbah, ``Large language model based multi-objective optimization for integrated sensing and communication in UAV networks," \textit{IEEE Wireless Commun. Lett.}, vol. 14, no. 4, pp. 979--983, Apr. 2025.
\bibitem{isac1}
F. Liu, W. Yuan, C. Masouros, and J. Yuan, ``Radar-assisted predictive beamforming for vehicular links: Communication served by sensing,'' {\it{IEEE Trans. Wireless Commun.}},  vol. 19, no. 11, pp. 7704--7719, Nov. 2020.
\bibitem{isac2}
C. B. Barneto, S. D. Liyanaar achchi, M. Heino, T. Riihonen, and M. Valkama, ``Full duplex radio/radar technology: The enabler for advanced joint communication and sensing,'' {\it{IEEE Wireless Commun.}}, vol. 28, no. 1, pp. 82--88, Feb. 2021.
\bibitem{Zhao}
D. Zhao, H. Lu, Y. Wang, H. Sun, and Y. Gui, ``Joint power allocation and user association optimization for IRS-assisted mmWave systems,'' {\it{IEEE Wireless Commun.}}, vol. 21, no. 1, pp. 577--590, Jan. 2022.
\bibitem{Astely}
D. Astely, E. Dahlman, A. Furuskr, Y. Jading, M. Lindstrm, and
S. Parkvall, “LTE: the Evolution of Mobile Broadband,” {\it{IEEE Commun. Mag.}}, vol. 47, no. 4, pp. 44–51, Apr. 2009.
\bibitem{Bekkerman} I. Bekkerman and J. Tabrikian, ``Target detection and localization using MIMO radars and sonars,'' {\it{IEEE Trans. Signal Process.}}, vol. 54, no. 10, pp. 3873--3883, Oct. 2006.
\bibitem{book3}
T. L. Marzetta, \textit{Fundamentals of Massive MIMO}. Cambridge, U.K.: Cambridge Univ. Press, 2016.
\bibitem{book2} S.  M.  Kay, {\it{Fundamentals of Statistical Signal Processing: Detection Theory}}, vol. 1. Englewood Cliffs, NJ, USA: Prentice--Hall, 1998.
\bibitem{LLM1}
T. B. Brown et al., ``Language models are few-shot learners,'' in \textit{ Proc. Adv. Neural Inf. Process. Syst.}, 2020, pp. 1877--1901.
\bibitem{LLM2}
J. Wei, et al., ``Chain of thought prompting elicits reasoning in large
language models,'' \textit{in Proc. Adv. Neural Inf. Process. Syst.}, 2022,
vol. 35, pp. 24824--24837.
\bibitem{LLM3}
N. Shinn, F. Cassano, B. Labash, A. Gopinath, K. Narasimhan, and S. Yao, ``Reflexion: Language agents with verbal reinforcement learning,'' 2023. [Online]. Available: https://arxiv.org/abs/2303.11366.
\bibitem{FP1}
K. Shen and W. Yu, ``Fractional programming for communication systems--Part I: Power control and beamforming,'' \textit{IEEE Trans. Signal Process.}, vol. 66, no. 10, pp. 2616--2630, May 2018.
\bibitem{FP2}
K. Shen and W. Yu, ``Fractional programming for communication systems--Part \Rmnum{2}: Uplink scheduling via matching,'' \textit{IEEE Trans. Signal Process.}, vol. 66, no. 10, pp. 2631--2644, May 2018.
\bibitem{wang}
Z. Wang, X. Mu, and Y. Liu, ``STARS enabled integrated sensing and communication,'' \textit{IEEE Trans. Wireless Commun.}, vol. 22, no. 10, pp. 6750--6765, Oct. 2023.
\bibitem{AL}
R. Andreani, E. G. Birgin, J. M. Martinez, and M. L. Schuverdt, ``On
augmented Lagrangian methods with general lower-level constraints,''
\textit{SIAM J. Optim.}, vol. 18, no. 4, pp. 1286--1309, Nov. 2007.
\bibitem{cvx}
M. Grant and S. Boyd, ``CVX: Matlab software for disciplined convex
programming, version 2.1,'' http://cvxr.com/cvx, Mar. 2014.
\bibitem{MM1}
Y. Sun, P. Babu, and D. P. Palomar, ``Majorization--minimization algorithms in signal processing, communications, and machine learning,'' \textit{IEEE Trans. Signal Process.}, vol. 65, no. 3, pp. 794--816, Feb. 2017.
\bibitem{gale-shapley}
L. E. Dubins and D. A. Freedman, ``Machiavelli and the gale-shapley
algorithm,'' \textit{The American Mathematical Monthly}, vol. 88, no. 7, pp.
485--494, 1981.
\bibitem{coalition}
M. Sami and J. N. Daigle, ``User association and power control for UAV-enabled cellular networks,'' \textit{IEEE Wireless Commun. Lett.}, vol. 9, no. 3, pp. 267--270, Mar. 2020.
\bibliographystyle{IEEEbib}
\end{thebibliography}
\end{document}